\documentclass[10pt, prd, aps, showpacs, nofootinbib]{revtex4-1}
\usepackage{geometry} 
\usepackage{amsmath,amssymb}
\usepackage{graphicx}
\usepackage{subfig}
\usepackage{caption}
\newcommand{\be}{\begin{equation}}
\newcommand{\ee}{\end{equation}}
\newcommand{\bea}{\begin{eqnarray}}
\newcommand{\eea}{\end{eqnarray}}

\newcommand{\Romatre}{Dip. di Matematica e Fisica, Universit\`a  Roma Tre and INFN, Sezione di Roma Tre,\\ Via della Vasca Navale 84, I-00146 Rome, Italy}
\newcommand{\RomatreINFN}{Istituto Nazionale di Fisica Nucleare, Sezione di Roma Tre,\\ Via della Vasca Navale 84, I-00146 Rome, Italy}

\begin{document}

\title{The light-quark contribution to the leading HVP term of the muon \\[2mm] $g - 2$ from twisted-mass fermions}

\author{D.~Giusti}\affiliation{\Romatre}
\author{F.~Sanfilippo}\affiliation{\RomatreINFN}
\author{S.~Simula}\affiliation{\RomatreINFN}

\begin{abstract}
\begin{center}
\vspace{0.5cm}
\includegraphics[draft=false,width=.25\linewidth]{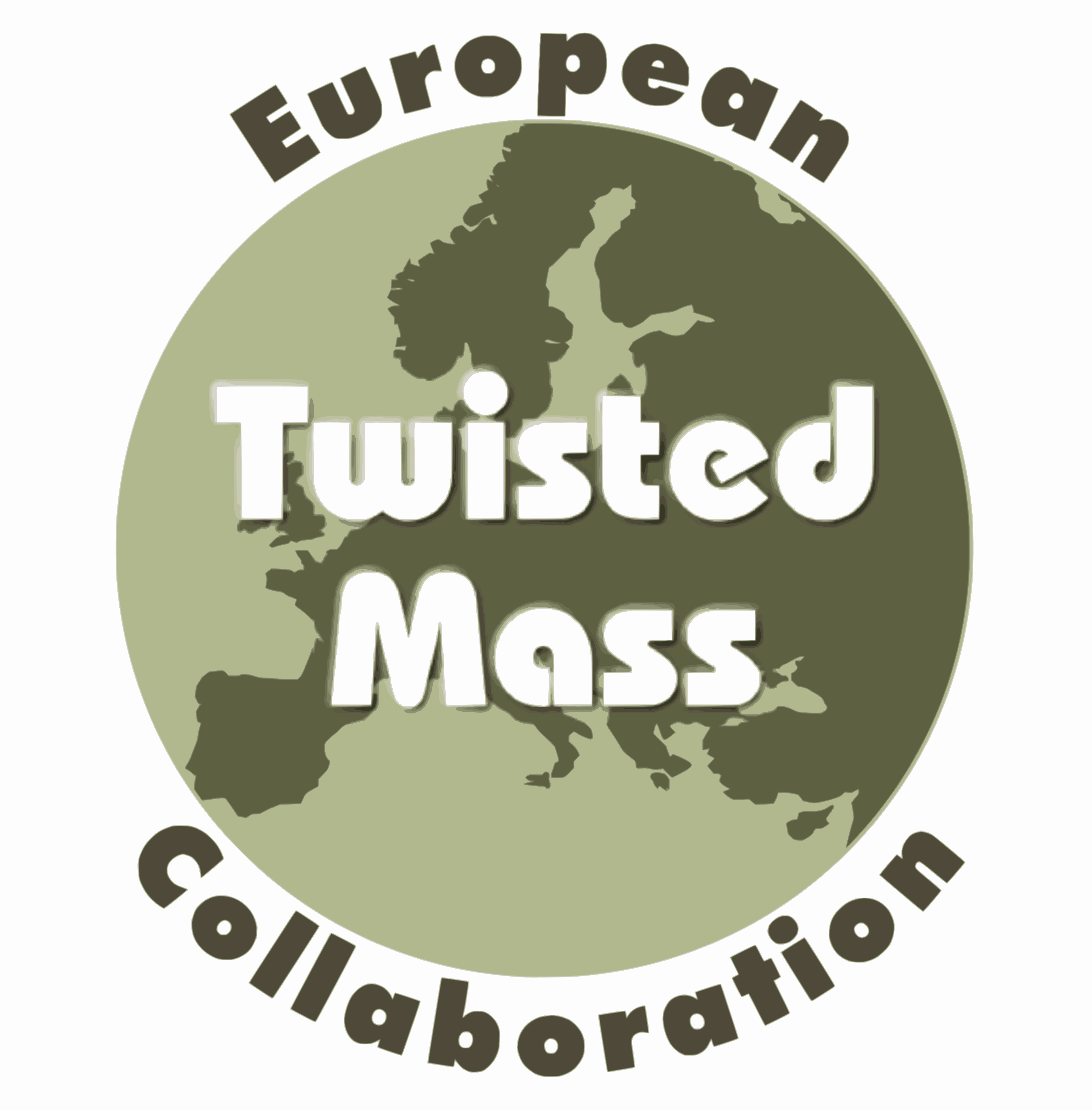}
\vspace{0.5cm}
\end{center}
We present a lattice calculation of the leading Hadronic Vacuum Polarization (HVP) contribution of the light $u$- and $d$-quarks to the anomalous magnetic moment of the muon, $a_\mu^{\rm HVP}(ud)$, adopting the gauge configurations generated by the European Twisted Mass Collaboration (ETMC) with $N_f = 2+1+1$ dynamical quarks at three values of the lattice spacing ($a \simeq 0.062, 0.082, 0.089$ fm) with pion masses in the range $M_\pi \simeq 210 - 450$ MeV. 
Thanks to several lattices at fixed values of the light-quark mass and scale but with different sizes we perform a careful investigation of finite-volume effects (FVEs), which represent one of main source of uncertainty in modern lattice calculations of $a_\mu^{\rm HVP}(ud)$.
In order to remove FVEs we develop an analytic representation of the vector correlator, which describes the lattice data for time distances larger than $\simeq 0.2$ fm.
The representation is based on quark-hadron duality at small and intermediate time distances and on the two-pion contributions in a finite box at larger time distances.
After removing FVEs we extrapolate the corrected lattice data to the physical pion point and to the continuum limit taking into account the chiral logs predicted by Chiral Perturbation Theory (ChPT). We obtain $a_\mu^{\rm HVP}(ud) = 619.0 ~ (17.8) \cdot 10^{-10}$. Adding the contribution of strange and charm quarks, obtained by ETMC, and an estimate of the isospin-breaking corrections and quark-disconnected diagrams from the literature we get $a_\mu^{\rm HVP}(udsc) = 683 ~ (19) \cdot 10^{-10}$, which is consistent with recent results based on dispersive analyses of the experimental cross section data for $e^+ e^-$ annihilation into hadrons. 
Using our analytic representation of the vector correlator, taken at the physical pion mass in the continuum and infinite volume limits, we provide the first eleven moments of the polarization function and we compare them with recent results of the dispersive analysis of the $\pi^+ \pi^-$ channels.
We estimate also the light-quark contribution to the missing part of $a_\mu^{\rm HVP}$ not covered in the MUonE experiment.
\end{abstract}

\maketitle

\newpage

\section{Introduction}
\label{sec:intro}

The anomalous magnetic moment of the muon $a_\mu \equiv (g - 2) / 2$ is one of the most precisely determined quantities in particle physics.
It is known experimentally with an accuracy of $0.54$ ppm~\cite{Bennett:2006fi} (BNL E821) and the current precision of the Standard Model (SM) prediction is at the level of $0.4$ ppm~\cite{PDG}.
The tension between the experimental value $a_\mu^{exp}$ and the SM prediction $a_\mu^{SM}$ corresponds to $\simeq 3.5 \div 4$ standard deviations, according to the most recent determinations of the Hadronic Vacuum Polarization (HVP) contribution, namely
 \bea
     a_\mu^{exp} - a_\mu^{SM} & = & 31.3 ~ (4.9)_{\rm th} ~ (6.3)_{\rm exp} ~ [7.7] \cdot 10^{-10} \qquad \mbox{\cite{Jegerlehner:2017lbd}} ~ , \nonumber \\
                                                 & = & 26.8 ~ (4.3)_{\rm th} ~ (6.3)_{\rm exp} ~ [7.6] \cdot 10^{-10} \qquad \mbox{\cite{Davier:2017zfy}} ~ , \nonumber \\
                                                 & = & 27.1 ~ (3.6)_{\rm th} ~ (6.3)_{\rm exp} ~ [7.3] \cdot 10^{-10} \qquad \mbox{\cite{Keshavarzi:2018mgv}} ~ ,
     \label{eq:tension_SM}
\eea     
where the first error is from the SM theory (mainly the HVP term), the second one from the experiment and the third one corresponds to their sum in quadrature. 

Since the tension given in Eq.~(\ref{eq:tension_SM}) might be an exciting indication of new physics (NP) beyond the SM, an improvement of the uncertainties is highly desirable.
The forthcoming $g - 2$ experiments at Fermilab (E989)~\cite{Logashenko:2015xab} and J-PARC (E34)~\cite{Otani:2015lra} aim at reducing the experimental uncertainty by a factor of four, down to 0.14 ppm, making the comparison of the experimental value of $a_\mu$ with the theoretical predictions one of the most important tests of the SM in the quest of NP effects.
With such a reduced experimental error, the uncertainty of the hadronic corrections, due to the HVP and hadronic light-by-light (LBL) terms \cite{Jegerlehner:2009ry}, will soon become the main limitation of this SM test. 

The theoretical predictions for the hadronic contribution $a_\mu^{\rm HVP}$ have been traditionally obtained from experimental data using dispersion relations for relating the HVP function to the experimental cross section data for $e^+ e^-$ annihilation into hadrons \cite{Davier:2010nc,Hagiwara:2011af}. 
An alternative approach was proposed in Refs.~\cite{Lautrup:1971jf,deRafael:1993za,Blum:2002ii}, namely to compute $a_\mu^{\rm HVP}$ in lattice QCD from the Euclidean correlation function of two electromagnetic (em) currents. 
In this respect an impressive progress in the lattice determinations of $a_\mu^{\rm HVP}$ has been achieved in the last few years \cite{Boyle:2011hu,DellaMorte:2011aa,Burger:2013jya,Chakraborty:2014mwa,Chakraborty:2015cso,Bali:2015msa,Chakraborty:2015ugp,Blum:2015you,Blum:2016xpd,Chakraborty:2016mwy,DellaMorte:2017dyu,Borsanyi:2017zdw,Blum:2018mom} and very interesting attempts to compute also the LBL contribution are under way both on the lattice \cite{Green:2015sra,Blum:2015gfa} and via dispersion approaches and Chiral Perturbation Theory (ChPT) \cite{Colangelo:2015ama,Bijnens:2016hgx,Colangelo:2017fiz}.

An updated status of lattice (as well as nonlattice) efforts for evaluating the hadronic corrections to $a_\mu$ can be found in Ref.~\cite{gm2_TI}.
The main open issue concerning the most accurate lattice calculations of $a_\mu^{\rm HVP}$, performed using gauge configurations at the physical pion point, is a significative tension between the HPQCD~\cite{Chakraborty:2016mwy} result, $a_\mu^{\rm HVP} = 667 (13) \cdot 10^{-10}$, on one hand side and the BMW~\cite{Borsanyi:2017zdw} and RBC/UKQCD~\cite{Blum:2018mom} findings, $a_\mu^{\rm HVP} = 711.0 (18.9) \cdot 10^{-10}$ and $a_\mu^{\rm HVP} = 715.4  (18.7) \cdot 10^{-10}$ respectively, on the other hand side. 
Such a tension originates almost totally from the light u- and d-quark (connected) contribution to the HVP and it turns out to be at the same level of the muon anomaly (\ref{eq:tension_SM}).

Besides the leading HVP correction to the one-loop muon diagram, which is of order ${\cal{O}}(\alpha_{em}^2)$, the increasing precision of the lattice calculations makes it necessary to include both em and strong isospin-breaking (IB) corrections, which contribute at order $O(\alpha_{em}^3)$ and $O(\alpha_{em}^2 (m_d - m_u))$ to the HVP, respectively. 
In Ref.~\cite{Giusti:2017jof} a lattice calculation of both the leading and the IB corrections to the HVP contribution due to strange and charm quark intermediate states was carried out using the time-momentum representation for $a_\mu^{\rm HVP}$ \cite{Bernecker:2011gh} and the expansion method of the path integral in the small parameters $\alpha_{em}$ and $(m_d - m_u) / \Lambda_{QCD}$~\cite{deDivitiis:2011eh,deDivitiis:2013xla}.
In the strange and charm sectors the strong IB corrections are absent at leading order in $(m_d - m_u)$, while the em corrections have been found to be negligible with respect to present uncertainties.
Other recent calculations of the IB corrections to the HVP have been performed in Refs.~\cite{Boyle:2017gzv,Chakraborty:2017tqp,Blum:2018mom}, while higher-order corrections due to diagrams containing HVP and lepton insertions have been recently estimated on the lattice in Ref.~\cite{Chakraborty:2018iyb}.

In this paper we present the results of a new lattice calculation of the leading HVP contribution due to light u- and d-quark (connected) intermediate states, $a_\mu^{\rm HVP}(ud)$, while the evaluation of the corresponding IB corrections will be addressed in a separate work.
We make use of the gauge ensembles generated by the European Twisted Mass Collaboration (ETMC) with $N_f = 2 + 1 + 1$ dynamical quarks, which include in the sea, besides two light mass-degenerate quarks, also the strange and the charm quarks with masses close to their physical values~\cite{Baron:2010bv,Baron:2011sf}.

Thanks to the various lattice volumes of the ETMC gauge ensembles we observe quite relevant finite volume effects (FVEs) for $a_\mu^{\rm HVP}(ud)$. 
Thus, we develop an analytic representation of the temporal dependence of the Euclidean vector correlator, based on the quark-hadron duality \cite{SVZ}, already observed in Ref.~\cite{Giusti:2017jof}, and on the two-pion contributions in a finite box \cite{Luscher:1985dn,Luscher:1986pf,Luscher:1990ux,Luscher:1991cf,Lellouch:2000pv,Meyer:2011um,Francis:2013qna}.
Using such a representation, which constitutes the original part of this work, we are able to reproduce accurately the temporal dependence of the Euclidean vector correlator for all the ETMC gauge ensembles and, by taking properly the infinite volume limit, we can correct in a systematic way our lattice values of $a_\mu^{\rm HVP}(ud)$ for the FVEs.
We point out that our estimate of FVEs takes into account the resonant interaction in the two-pion system at variance with the ChPT prediction at next-to-leading (NLO) order~\cite{Aubin:2015rzx}.

The main result of the present study is
 \be
       a_\mu^{\rm HVP}(ud) = 619.0 ~ (14.7)_{stat+fit+input} (6.2)_{chir} (4.9)_{disc} (6.2)_{FVE} ~ [17.8] \cdot 10^{-10} ~ ,
       \label{eq:result_ud}
 \ee
where the errors come in the order from (statistics + fitting procedure + input parameters), chiral extrapolation, discretization and finite volume effects.

Our result (\ref{eq:result_ud}) improves the previous ETMC estimate of Ref.~\cite{Burger:2013jya} and agrees within the errors with the HPQCD $a_\mu^{\rm HVP}(ud) = 599 ~ (11) \cdot 10^{-10}$~\cite{Chakraborty:2016mwy}, the CLS/Mainz $a_\mu^{\rm HVP}(ud) = 588.2 ~ (35.8) \cdot 10^{-10}$~\cite{DellaMorte:2017dyu}, the BMW $a_\mu^{\rm HVP}(ud) = 647.5 ~ (19.2) \cdot 10^{-10}$~\cite{Borsanyi:2017zdw} and the RBC/UKQCD $a_\mu^{\rm HVP}(ud) = 649.7 ~ (15.0) \cdot 10^{-10}$~\cite{Blum:2018mom} results. 

Adding the (connected) contributions from strange and charm quarks, $a_\mu^{\rm HVP}(s) = 53.1 ~ (2.5) \cdot 10^{-10}$ and $a_\mu^{\rm HVP}(c) = 14.75 ~ (0.56) \cdot 10^{-10}$ determined by ETMC in Ref.~\cite{Giusti:2017jof}, and an estimate of the IB corrections $a_\mu^{\rm HVP}(IB) = 8 ~ (5) \cdot 10^{-10}$ and of the quark disconnected diagrams $a_\mu^{\rm HVP}(disconn.) = -12 ~ (4) \cdot 10^{-10}$, obtained using the findings of Refs.~\cite{Borsanyi:2017zdw} and \cite{Blum:2018mom}, we finally get
 \be
       a_\mu^{\rm HVP}(udsc) = 683 ~ (19) \cdot 10^{-10} ~ ,
       \label{eq:result_full}
 \ee
which is in nice agreement with the recent  results $a_\mu^{\rm HVP} = 688.07 ~ (4.14) \cdot 10^{-10}$~\cite{Jegerlehner:2017lbd}, $a_\mu^{\rm HVP} = 693.10 ~ (3.40) \cdot 10^{-10}$~\cite{Davier:2017zfy} and $a_\mu^{\rm HVP} = 693.27 ~ (2.46) \cdot 10^{-10}$~\cite{Keshavarzi:2018mgv}, based on dispersive analyses of the experimental cross section data for $e^+ e^-$ annihilation into hadrons. 

Using our analytic representation of the vector correlator, taken at the physical pion mass in the continuum and infinite volume limits, we provide the slope and curvature of the polarization function, $ \Pi_1^{(ud)} = 0.1642 ~ (33) ~ \mbox{GeV}^{-2}$ and $\Pi_2^{(ud)} = - 0.383 ~ (16) ~ \mbox{GeV}^{-4}$, which are compared with lattice results available in the literature.
We also estimate higher-order moments (up to the eleventh moment) and compare them with the values of the dispersive analysis of the $\pi^+ \pi^-$ channels made in Ref.~\cite{Keshavarzi:2018mgv}.
Finally, we estimate the light-quark contribution to the missing part of $a_\mu^{\rm HVP}$ not covered in the MUonE experiment~\cite{Calame:2015fva,Abbiendi:2016xup}.

The paper is organized as follows. 
In section~\ref{sec:TMR} we introduce the basic quantities and notation.
After providing the simulation details and addressing the identification of the ground-state, we evaluate $a_\mu^{\rm HVP}(ud)$ for all the ETMC ensembles and show the relevance of FVEs.  
In section~\ref{sec:representation} we develop an analytic representation of the vector correlator, based on the quark-hadron duality and the two-pion contributions, obtaining a quite accurate reproduction of the lattice data of the light-quark vector correlator.
In section~\ref{sec:FVE} we remove FVEs from the lattice data using the analytic representation, while in section~\ref{sec:extrapolations} we perform the extrapolations to the physical pion point and to the continuum limit.
Our findings are then compared with lattice results available in the literature.
In section~\ref{sec:physical} we discuss some relevant features of the analytic representation extrapolated at the physical pion mass and in the continuum limit. 
We provide the estimates of the lowest-order moments of the polarization function and compare them with the lattice results available in the literature. 
We estimate also higher-order moments, which we compare with the values of the dispersive analysis of the $\pi^+ \pi^-$ channels made in Ref.~\cite{Keshavarzi:2018mgv}, as well as the light-quark contribution to the missing part of $a_\mu^{\rm HVP}$ not covered in the MUonE experiment~\cite{Calame:2015fva,Abbiendi:2016xup}.
Finally, section~\ref{sec:conclusions} contains our conclusions and outlooks for future developments.

\section{Time-momentum representation}
\label{sec:TMR}

Following our previous work~\cite{Giusti:2017jof}, we adopt the time-momentum representation for the evaluation of $a_\mu^{\rm HVP}$, namely
 \be
      a_\mu^{\rm HVP} = 4 \alpha_{em}^2 \int_0^\infty dt ~ f(t) V(t) ~ ,
      \label{eq:amu_t}
 \ee
 where $t$ is the Euclidean time, the kernel function $f(t)$ is given by \cite{Bernecker:2011gh}
  \be
      f(t) \equiv \frac{4}{m_\mu^2} \int_0^\infty dz ~ \frac{1}{\sqrt{4 + z^2}} ~ \left( \frac{\sqrt{4 + z^2} - z}{\sqrt{4 +z^2} +z} \right)^2 
                       \left[ \frac{\mbox{cos}(z\,m_\mu t) -1}{z^2} + \frac{1}{2} m_\mu^2 t^2 \right] 
      \label{eq:ftilde}
  \ee
and the (Euclidean) vector correlator $V(t)$ is defined as
 \be
     V(t) \equiv \frac{1}{3} \sum_{i=x,y,z} \int d\vec{x} ~ \langle J_i(\vec{x}, t) J_i(0) \rangle
     \label{eq:VV}
 \ee
with $J_\mu(x)$ being the em current operator
 \be
     J_\mu(x) \equiv \sum_{f = u, d, s, c, ...} q_f ~ \overline{\psi}_f(x) \gamma_\mu \psi_f(x) ~ .
     \label{eq:Jmu}
 \ee 

The vector correlator $V(t)$ can be calculated on a lattice with volume $L^3$ and temporal extension $T$ at discretized values of the time distance $t$ from $0$ to $T$. 
In this work we will limit ourselves to the contribution of the light $u$- and $d$-quarks, evaluated in isosymmetric QCD ($m_u = m_d = m_{ud}$) neglecting also off-diagonal flavor terms (i.e., including quark-connected diagrams only).
Thus, one gets
\be
     a_\mu^{\rm HVP}(ud) = 4 \alpha_{em}^2 \Bigl\{ \sum_{t = 0}^{T_{data}} f(t) V^{(ud)}(t) + \sum_{t = T_{data} + a}^\infty f(t) \frac{Z_V^{(ud)}}{2 M_V^{(ud)}} 
                                           e^{- M_V^{(ud)} t} \Bigr\} ~ ,
     \label{eq:decomposition}
\ee
where the first term in the r.h.s~is directly given by the lattice data, while for the second term the identification of the ground-state at large time distances is required (see Refs.~\cite{Chakraborty:2015ugp,Blum:2015you,Chakraborty:2016mwy,DellaMorte:2017dyu,Giusti:2017jof}).
In Eq.~(\ref{eq:decomposition}) $M_V^{(ud)}$ is the ground-state mass and $Z_V^{(ud)}$ is the squared matrix element of the light-quark vector current between the vacuum and the state $| V \rangle$: $Z_V^{(ud)} \equiv (1/3) \sum_{i=x,y,z}$ $\sum_{f=u,d} q_f^2$ $| \langle 0 | \overline{\psi}_f(0) \gamma_i \psi_f(0) | V \rangle |^2$. 
The value of $T_{data}$ has to be large enough that the ground-state contribution is dominant for $t > T_{data}$ and smaller than $T / 2$ in order to avoid backward signals.
An important consistency check is that the sum of the two terms in the r.h.s.~of Eq.~(\ref{eq:decomposition}) should be independent of the specific choice of the value of $T_{data}$, as it will be shown later in section \ref{sec:GS}.

\subsection{Simulation details}
\label{sec:simulations}

The gauge ensembles used in this work are the same adopted in Ref.~\cite{Carrasco:2014cwa} to determine the up, down, strange, charm quark masses and the lattice scale. 
We employ the Iwasaki action \cite{Iwasaki:1985we} for gluons and the Wilson Twisted Mass Action \cite{Frezzotti:2000nk, Frezzotti:2003xj, Frezzotti:2003ni} for sea quarks.

We have considered three values of the inverse bare lattice coupling $\beta$ and different lattice volumes, as shown in Table~\ref{tab:simudetails}, where the number of configurations analyzed ($N_{cfg}$) corresponds to a separation of $20$ trajectories.
For earlier investigations of FVEs ETMC had produced three dedicated ensembles, A40.20, A40.24 and A40.32, which share the same quark mass and lattice spacing and differ only in the lattice size $L$.
To improve such an investigation, which is crucial in the present work, a further gauge ensemble, A40.40, has been generated at a larger value of the lattice size $L$.

We work in isosymmetric QCD ($m_u = m_d = m_{ud}$) and at each lattice spacing different values of the light sea quark masses have been considered. 
The light valence and sea quark masses are always taken to be degenerate ($m_{ud}^{sea} = m_{ud}^{val} = m_{ud}$). 

\begin{table}[hbt!]
{\small
\begin{center}
\begin{tabular}{||c||c|c||c|c|c||c||c|c|c||}
\hline
ensemble & $\beta$ & $V / a^4$ &$a\mu_{ud}$&$a\mu_\sigma$&$a\mu_\delta$&$N_{cfg}$& $L~(\mbox{fm})$ & $M_\pi~(\mbox{MeV})$ & $M_\pi L$  \\
\hline \hline
$A40.40$ & $1.90$ & $40^{3}\times 80$ &$0.0040$ &$0.15$ &$0.19$ & $100$ & 3.5 & 317 (12) & 5.7 \\
\cline{1-1} \cline{3-4} \cline{7-10}
$A30.32$ & & $32^{3}\times 64$ &$0.0030$ & & & $150$ & 2.8 & 275 (10) & 3.9 \\
$A40.32$ & & & $0.0040$ & & & $100$ & & 316 (12) & 4.5  \\
$A50.32$ & & & $0.0050$ & & & $150$ & & 350 (13) & 5.0  \\
\cline{1-1} \cline{3-4} \cline{7-10}
$A40.24$ & & $24^{3}\times 48 $ & $0.0040$ & & & $150$ & 2.1 & 322 (13) & 3.5 \\
$A60.24$ & & & $0.0060$ & & & $150$ & & 386 (15) & 4.2 \\
$A80.24$ & & & $0.0080$ & & & $150$ & & 442 (17) & 4.8 \\
$A100.24$ &  & & $0.0100$ & & & $150$ & & 495 (19) & 5.3 \\
\cline{1-1} \cline{3-4} \cline{7-10}
$A40.20$ & & $20^{3}\times 48 $ & $0.0040$ & & & $150$ & 1.8 & 330 (13) & 3.0 \\
\hline \hline
$B25.32$ & $1.95$ & $32^{3}\times 64$ &$0.0025$&$0.135$ &$0.170$& $150$ & 2.6 & 259 (9) & 3.4 \\
$B35.32$ & & & $0.0035$ & & & $150$ & & 302 (10) & 4.0 \\
$B55.32$ & & & $0.0055$ & & & $150$ & & 375 (13) & 5.0 \\
$B75.32$ &  & & $0.0075$ & & & $~80$ & & 436 (15) & 5.8 \\
\cline{1-1} \cline{3-4} \cline{7-10}
$B85.24$ & & $24^{3}\times 48 $ & $0.0085$ & & & $150$ & 2.0 & 468 (16) & 4.6 \\
\hline \hline
$D15.48$ & $2.10$ & $48^{3}\times 96$ &$0.0015$&$0.1200$ &$0.1385 $& $100$& 3.0 & 223 (6) & 3.4 \\ 
$D20.48$ & & & $0.0020$ & & & $100$ & & 256 (7) & 3.0 \\
$D30.48$ & & & $0.0030$ & & & $100$ & & 312 (8) & 4.7 \\
 \hline   
\end{tabular}
\end{center}
}
\vspace{-0.25cm}
\caption{\it \small Values of the simulated quark bare masses (in lattice units), of the pion mass $M_\pi$, of the lattice size $L$ and of the product $M_\pi L$ for the $16$ ETMC gauge ensembles with $N_f = 2+1+1$ dynamical quarks used in this work (see Ref.~\cite{Carrasco:2014cwa}) and for the gauge ensemble, A40.40 added to improve the investigation of FVEs. The bare twisted masses $\mu_\sigma$ and $\mu_\delta$ describe the strange and charm sea doublet according to Ref.~\cite{Frezzotti:2003xj}. The central values and errors of the pion mass  are evaluated using the bootstrap events of the eight branches of the analysis of Ref.~\cite{Carrasco:2014cwa}. The valence quarks in the pion are regularized with opposite values of the Wilson $r$-parameter in order to guarantee that discretization effects on the pion mass are of order $\mathcal{O}(a^2 \mu_{ud} ~ \Lambda_{QCD})$.\hspace*{\fill}}
\label{tab:simudetails}
\end{table}

In this work we made use of the bootstrap samplings elaborated for the input parameters of the quark mass analysis of Ref.~\cite{Carrasco:2014cwa}.
There, eight branches of the analysis were adopted differing in: 
\begin{itemize}
\item the continuum extrapolation adopting for the scale parameter either the Sommer parameter $r_0$ or the mass of a fictitious pseudoscalar meson made up of strange(charm)-like quarks; 
\item the chiral extrapolation performed with fitting functions chosen to be either a polynomial expansion or a ChPT Ansatz in the light-quark mass;
\item the choice between the methods M1 and M2, which differ by $O(a^2)$ effects, used to determine the mass renormalization constant (RC) $Z_m = 1 / Z_P$ in the RI'-MOM scheme. 
\end{itemize}

Throughout this work the renormalized light-quark mass $m_{ud}$ is given in the $\overline{MS}$ scheme at a renormalization scale equal to $2$ GeV.
At the physical pion point ($M_\pi^{phys} = M_{\pi^0} = 135$ MeV) the value $m_{ud}^{phys} = 3.70 ~ (17)$ MeV was determined in Ref.~\cite{Carrasco:2014cwa}, using  the experimental value of the pion decay constant for fixing the lattice scale.

\subsection{Ground-state identification}
\label{sec:GS}

As in Ref.~\cite{Giusti:2017jof}, in the numerical simulations we have adopted the following local version of the vector current
 \be
    J_\mu(x) = Z_A ~ \bar{\psi}^\prime(x) \gamma_\mu \psi(x) ~ ,
    \label{eq:localV}
 \ee
where $\psi^\prime$ has the same mass and charge of $\psi$, but it is regularized with an opposite value of the Wilson $r$-parameter, i.e.~$r^\prime = - r$.
Being at maximal twist the current (\ref{eq:localV}) renormalizes multiplicatively through the RC $Z_A$ determined in Ref.~\cite{Carrasco:2014cwa}.
By construction the local current (\ref{eq:localV}) cannot generate off-diagonal flavor contributions in the vector correlator (\ref{eq:VV}).

As discussed in Ref.~\cite{Giusti:2017jof}, the properties of  the kernel function $f(t)$, given by Eq.~(\ref{eq:ftilde}), guarantee that the contact terms, generated in the HVP tensor by a local vector current, cannot contribute to the evaluation of $a_\mu^{\rm HVP}$ (see also Ref.~\cite{Burger:2014ada}).

We have calculated the vector correlator (\ref{eq:VV}) adopting the local current (\ref{eq:localV}) for the light $u$ and $d$-quarks using 160 stochastic sources (diagonal in the spin variable and dense in the color one) per gauge configuration.
For each gauge ensemble the ground-state mass $M_V^{(ud)}$ and the coupling constant $Z_V^{(ud)}$ are extracted from a single exponential fit (including the proper backward signal) in the range $t_{min} \leq t \leq t_{max}$.
The values chosen for $t_{min}$ and $t_{max}$ are collected in Table \ref{tab:GS_light}.
\begin{table}[hbt!]
\begin{center}
\renewcommand{\arraystretch}{1.10}
\begin{tabular}{||c|c||c|c||}
\hline
$\beta$ & $V / a^4$ & $t_{min} / a$ & $t_{max} / a$ \\
\hline \hline
$1.90$ & $40^3 \times 80$ &$12$ &$22$ \\
\cline{2-4}
            & $32^3 \times 64$ &$12$ &$22$ \\
\cline{2-4}
            & $24^3 \times 48$ &$12$ &$20$ \\
\cline{2-4}
            & $20^3 \times 48$ &$12$ &$20$ \\
\hline \hline
$1.95$ & $32^3 \times 64$ &$13$ &$22$ \\
\cline{2-4}
            & $24^3 \times 48$ & $13$ &$20$ \\
\hline \hline
$2.10$ & $48^3 \times 96$ & $18$ &$30$ \\ 
\hline   
\end{tabular}
\renewcommand{\arraystretch}{1.0}
\end{center}
\vspace{-0.25cm}
\caption{\it \small Values of $t_{min}$ and $t_{max}$ chosen to extract the ground-state signal from the light-quark vector correlator $V(t)$ for the ETMC gauge ensembles of Table~\ref{tab:simudetails}.\hspace*{\fill}}
\label{tab:GS_light}
\end{table}

\begin{figure}[htb!]
\centering{\scalebox{0.75}{\includegraphics{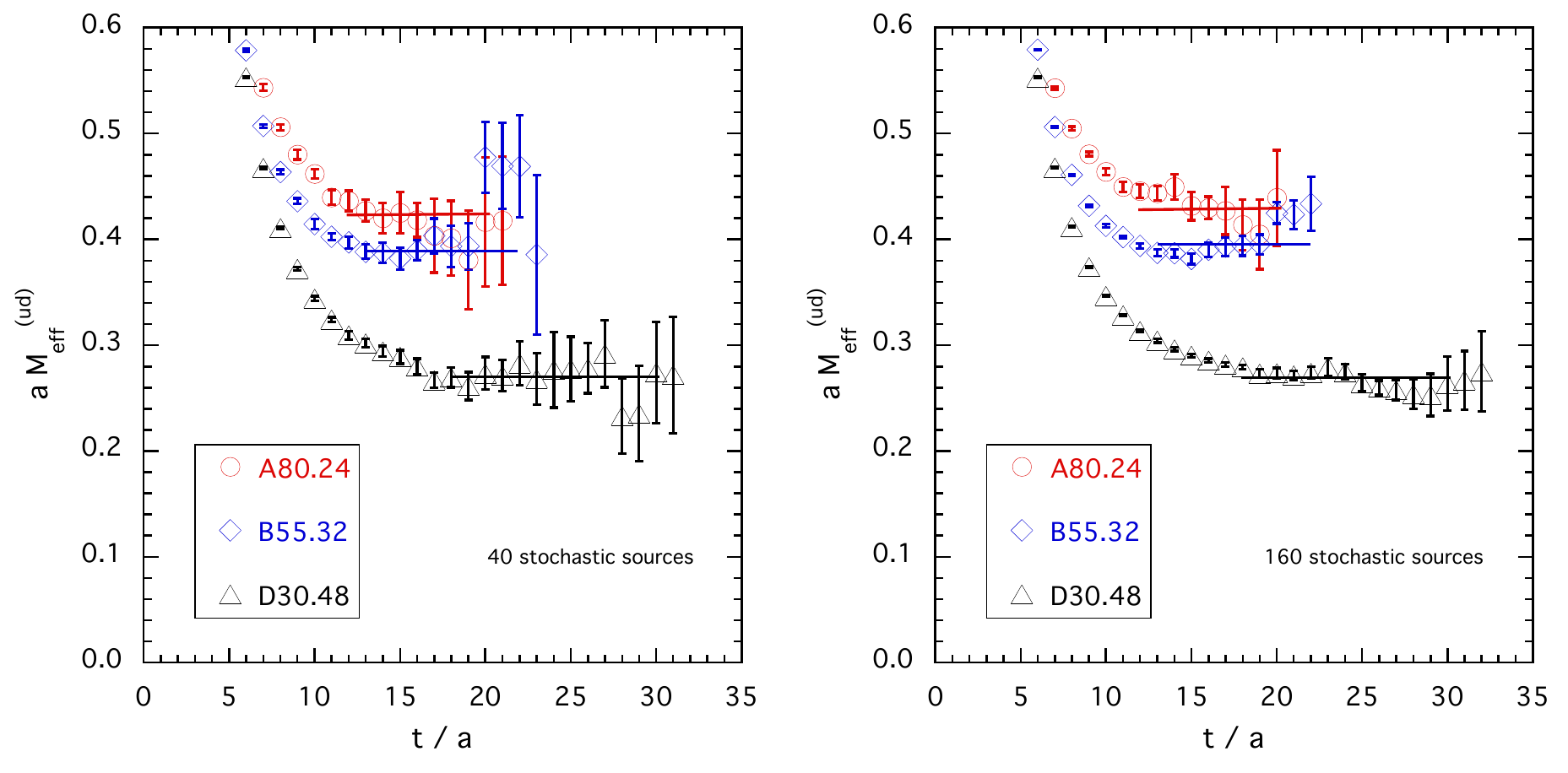}}}
\vspace{-0.5cm}
\caption{\it \small The effective mass $M_{eff}^{(ud)}(t)$ in lattice units (see Eq.~(\ref{eq:Meff})) corresponding to the light-quark vector correlator $V^{(ud)}(t)$ for the ETMC gauge ensembles A80.24, B55.32 and D30.48, evaluated using either 40 (left panel) or 160 (right panel) stochastic sources per each gauge configuration.\hspace*{\fill}}
\label{fig:Meff}
\end{figure}

The statistical precision of the effective mass $M_{eff}^{(ud)}(t)$, defined as
  \be
      a M_{eff}^{(ud)}(t) \equiv \mbox{arcosh}\left[ \frac{V^{(ud)}(t - a) + V^{(ud)}(t + a)} {2 V^{(ud)}(t)} \right] ~ _{\overrightarrow{t \geq t_{min}}} ~ a M_V^{(ud)} ~ ,
      \label{eq:Meff}
   \ee
is illustrated in Fig.~\ref{fig:Meff} by comparing the results obtained using either 40 or 160 stochastic sources per gauge configuration in the case of the ETMC ensembles A80.24, B55.32 and D30.48.
We observe that the increase of the number of stochastic sources is beneficial, but the quality of the plateaux at large time distances is nevertheless still limited.

\subsection{Lattice data and FVEs}
\label{sec:data}

We have evaluated Eq.~(\ref{eq:decomposition}) adopting four choices of $T_{data}$, namely: $T_{data} = (t_{min} + 2a)$, $(t_{min} + t_{max}) / 2$, $(t_{max} - 2a)$ and $(T / 2 - 4a)$, and using the values of the ground-state mass $M_V^{(ud)}$ and (squared) matrix elements $Z_V^{(ud)}$, determined, as described in the previous subsection, from a single exponential fit of the vector correlator $V^{(ud)}(t)$ in the range $t_{min} \leq t \leq t_{max}$, with the values of $t_{min}$ and $t_{max}$ given in Table \ref{tab:GS_light}.

The results obtained in the case of the ETMC gauge ensembles A40.24, B25.32 and D15.48 are collected in Table \ref{tab:Ndt_light} for illustrative purposes.
\begin{table}[hbt!]
\begin{center}
\centerline{ensemble A40.24} 
\vspace{0.15cm}
\begin{tabular}{||c||c|c|c|c||}
\hline
$a_\mu^{\rm HVP}(ud)$ & $ (t_{min} + 2a) $ & $ (t_{min} + t_{max}) / 2 $ &  $ (t_{max} - 2a) $ &  $ (T / 2 - 4a) $ \\
\hline \hline
$T \leq T_{data}$ & $274.4 ~~ (7.2)$ & $300.0 ~~ (7.4)$ & $319.3 ~~ (7.7)$ & $334.1 ~~ (9.0)$ \\
\hline
$T > T_{data}$     & $~78.7 ~ (10.0)$ & $~53.1 ~~ (8.4)$ & $~34.5 ~~ (6.6)$ & $~19.9 ~~ (4.6)$ \\
\hline
${\rm total}$         & $353.1 ~ (10.8)$ & $353.1 ~ (10.5)$ & $353.9 ~ (10.8)$ & $354.0 ~ (11.7)$ \\
\hline \hline
\end{tabular}

\vspace{0.25cm}
\centerline{ensemble B25.32} 
\vspace{0.15cm}

\begin{tabular}{||c||c|c|c|c||}
\hline
$a_\mu^{\rm HVP}(ud)$ & $ (t_{min} + 2a) $ & $ (t_{min} + t_{max}) / 2 $ &  $ (t_{max} - 2a) $ &  $ (T / 2 - 4a) $ \\
\hline \hline
$T \leq T_{data}$ & $289.1 ~~ (5.7)$ & $326.2 ~~ (7.3)$ & $360.3 ~ (9.4)$ & $395.3 ~ (14.7)$ \\
\hline
$T > T_{data}$     & $111.6 ~~ (9.8)$ & $~74.5 ~~ (8.0)$ & $~40.8 ~~ (5.5)$ & $~~6.6 ~~~ (1.4)$ \\
\hline
${\rm total}$         & $400.7 ~ (13.6)$ & $400.7 ~ (13.6)$ & $401.1 ~ (13.9)$ & $401.9 ~ (16.0)$ \\
\hline \hline
\end{tabular}

\vspace{0.25cm}
\centerline{ensemble D15.48} 
\vspace{0.15cm}

\begin{tabular}{||c||c|c|c|c||}
\hline
$a_\mu^{\rm HVP}(ud)$& $ (t_{min} + 2a) $ & $ (t_{min} + t_{max}) / 2 $ &  $ (t_{max} - 2a) $ &  $ (T / 2 - 4a) $ \\
\hline \hline
$T \leq T_{data}$ & $324.9 ~~ (6.3)$ & $380.8 ~~ (8.0)$ & $416.0 ~ (10.4)$ & $440.6 ~ (55.6)$ \\
\hline
$T > T_{data}$     & $133.4 ~ (12.6)$ & $~79.1 ~ (10.1)$ & $~41.6 ~~ (6.9)$ & $~~2.1 ~~~ (0.7)$ \\
\hline
${\rm total}$         & $458.3 ~ (15.1)$ & $459.9 ~ (15.1)$ & $457.6 ~ (15.7)$ & $442.7 ~ (55.7)$ \\
\hline \hline
\end{tabular}
\end{center}
\vspace{-0.25cm}
\caption{\it \small Results for the light-quark (connected) contribution to $a_\mu^{\rm HVP}(ud)$ in units of $10^{-10}$, obtained adopting in Eq.~(\ref{eq:decomposition}) $T_{data} = (t_{min} + 2a)$, $(t_{min} + t_{max}) / 2$, $(t_{max} - 2a)$ and $(T / 2 - 4a)$  for the ETMC gauge ensembles A40.24, B25.32 and D15.48.\hspace*{\fill}}
\label{tab:Ndt_light}
\end{table}
The two terms in the r.h.s.~of Eq.~(\ref{eq:decomposition}) depend on the specific value of $T_{data}$, as expected, but their total sum is almost independent of the specific choice of $T_{data}$.
In order to minimize the impact of the contribution depending on the identification of the ground-state signal and to optimize at the same time the statistical uncertainties the value $T_{data} = (t_{max} - 2a)$ has been adopted in what follows.

\begin{figure}[htb!]
\centering{\scalebox{0.75}{\includegraphics{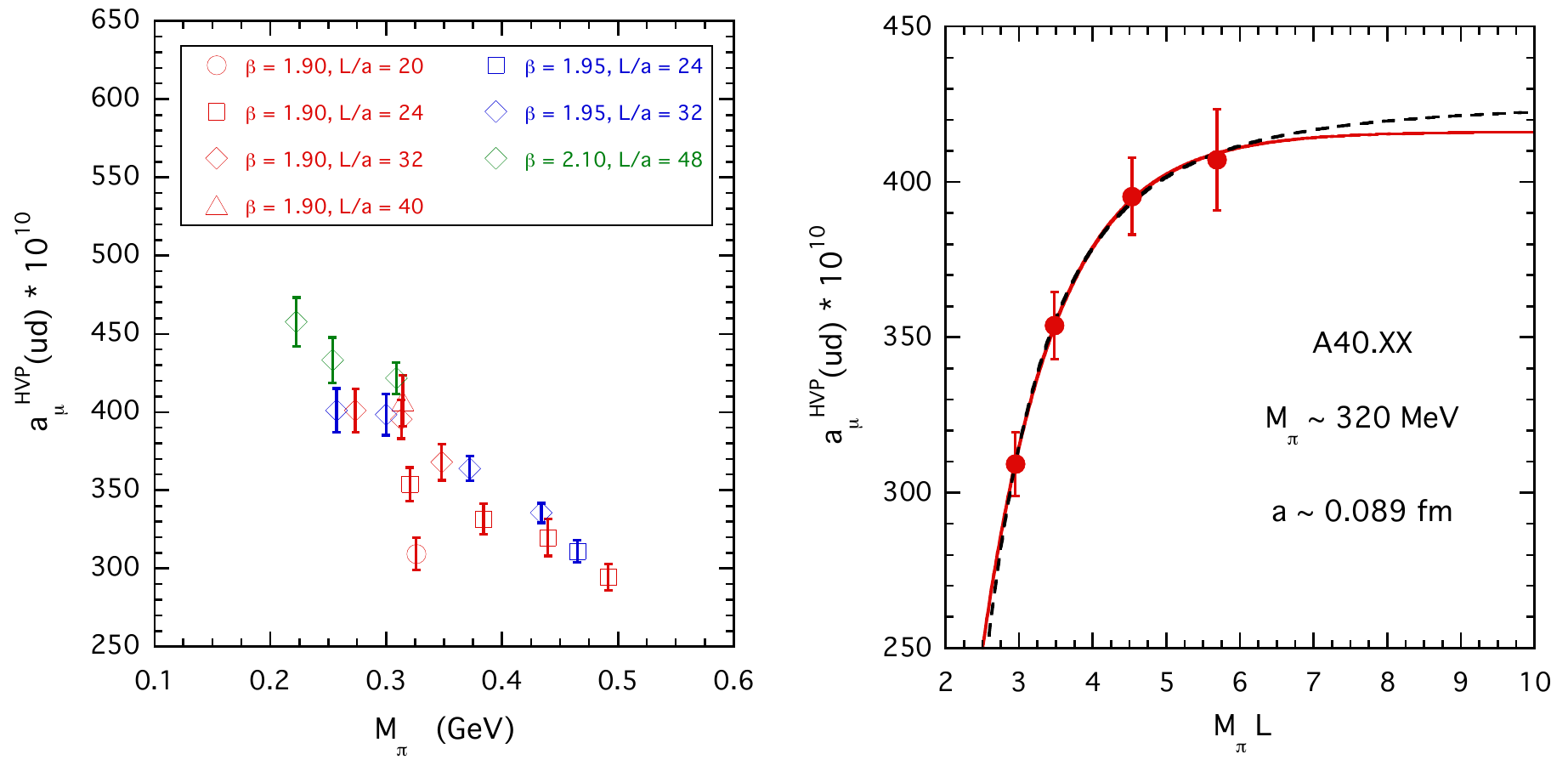}}}
\vspace{-0.25cm}
\caption{\it \small Left panel: results for $a_\mu^{\rm HVP}(ud)$ obtained using Eq.~(\ref{eq:decomposition}) (with $T_{data} = t_{max} - 2a$) for all the ETMC ensembles of Table \ref{tab:simudetails} versus the simulated pion mass $M_\pi$. Right panel: lattice data in the case of the four ensembles A40.XX with $XX = 20, 24, 32$ and $40$, corresponding to a pion mass $M_\pi \simeq 320$ MeV and a lattice spacing $a \simeq 0.089$ fm. The (red) solid and (black) dashed lines correspond, respectively, to an  exponential, $A (1 - B e^{-M_\pi L})$, and a power-law, $A^\prime (1 - B^\prime / (M_\pi L)^3)$, phenomenological fit.\hspace*{\fill}}
\label{fig:amul_ETMC}
\end{figure}

The results for $a_\mu^{\rm HVP}(ud)$ for all the ETMC ensembles of Table \ref{tab:simudetails} versus the simulated pion mass $M_\pi$ are collected in the left panel of Fig.~\ref{fig:amul_ETMC}, while the right panel contains only our findings in the case of the four ensembles A40.XX with $XX = 20, 24, 32$ and $40$, which share the same quark mass and lattice spacing and differ only in the lattice size $L$.

The lattice data for $a_\mu^{\rm HVP}(ud)$ exhibit a strong dependence on the pion mass and a remarkable sensitivity to FVEs at variance with the results obtained in the case of the strange and charm quark contributions to $a_\mu^{\rm HVP}$ (see Ref.~\cite{Giusti:2017jof}).
In particular, the data shown in the right panel of Fig.~\ref{fig:amul_ETMC} indicates that at a simulated pion mass $M_\pi \simeq 320$ MeV the FVEs are at the level of $\simeq 25\%$ for $M_\pi L \simeq 3$ and they reduce to $\simeq 5\%$ only at $M_\pi L \simeq 5$. 
The precision of the lattice data do not allow to distinguish whether the FVEs are exponentially or power-law suppressed~\cite{Luscher:1985dn,Luscher:1986pf}.

The large corrections observed for the ETMC ensembles A40.XX need to be understood and estimated properly. 
At NLO ChPT is unable to reproduce the value of $a_\mu^{\rm HVP}$~\cite{Aubin:2006xv} because of the important role of resonance contributions, which starts only at higher orders. 
The NLO chiral prediction for the FVEs is believed to be adequate close to the physical pion point~\cite{Aubin:2015rzx,Bijnens:2017esv}, since it is dominated by pion loops.
However, the NLO chiral result for the FVEs coincide with the estimate corresponding to noninteracting two-pion states in a finite box~\cite{Francis:2013qna,DellaMorte:2017dyu}. 
When applied at a pion mass of $\simeq 300$ MeV, we find that the NLO chiral prediction for FVEs is off by one order of magnitude with respect to what is observed in the right panel of Fig.~\ref{fig:amul_ETMC}. 
The $\rho$-meson resonant contribution to the interaction between two pions may therefore play an important role not only for $a_\mu^{\rm HVP}(ud)$, but also for the evaluation of FVEs.
Thus, we have elaborated an analytic representation of the vector correlator $V^{(ud)}(t)$, which incorporates resonant two-pion states and is given in terms of few quantities exhibiting small FVEs.
In this way we may achieve a good, direct control of FVEs in $a_\mu^{\rm HVP}(ud)$.
The analytic representation is described in the next section and the subtraction of FVEs is carried out in Section \ref{sec:FVE}.

\section{Analytic representation of the light-quark vector correlator}
\label{sec:representation}

In this section we develop an analytic representation of the temporal dependence of the vector correlator $V^{(ud)}(t)$, based on the quark-hadron duality~\cite{SVZ} and on the two-pion contributions in a finite box~\cite{Luscher:1985dn,Luscher:1986pf,Luscher:1990ux,Luscher:1991cf,Lellouch:2000pv,Meyer:2011um,Francis:2013qna}.

Let us start with the two-pion contribution, which in infinite volume is a continuous function above the two-particle threshold.
In a finite box of volume $L^3$ the two-pion states have been analyzed in detail in Refs.~\cite{Luscher:1985dn,Luscher:1986pf,Luscher:1990ux,Luscher:1991cf}.
The energy levels $\omega_n$ of the two-pion states are given by
 \be
     \omega_n = 2 \sqrt{M_\pi^2 + k_n^2} ~ ,
     \label{eq:omegan}
 \ee
where the discretized values $k_n$ should satisfy the L\"uscher condition, which for the case at hand (two pions in a $P$-wave with total isospin $1$) reads as 
 \be
     \delta_{11}(k_n) + \phi\left( \frac{k_nL}{2\pi} \right) = n \pi ~ ,
     \label{eq:kn}
 \ee
where $\delta_{11}$ is the (infinite volume) scattering phase shift and $\phi(z)$ is a known kinematical function defined as
\be
    \mbox{tan}\phi(z) = - \frac{2 \pi^2 z}{\sum_{\vec{m} \in \mathbb{Z}^3} \left( |\vec{m}|^2 - z^2 \right)^{-1}} ~ .
    \label{eq:phi}
\ee
The two-pion contribution to the vector correlator, $V_{\pi \pi}(t)$, can be written as~\cite{Lellouch:2000pv,Meyer:2011um,Francis:2013qna}
 \be
     V_{\pi \pi}(t) = \sum_n \nu_n |A_n|^2 e^{-\omega_n t} ~ ,
     \label{eq:V2pi}
 \ee
where $\nu_n$ is the number of vectors $\vec{z} \in \mathbb{Z}^3$ with norm $|\vec{z}|^2 = n$ and the squared amplitudes $|A_n|^2$ are related to the square of the timelike pion form factor $|F_\pi(\omega)|^2$ by
 \be
     \nu_n |A_n|^2 = \frac{2 k_n^5}{3 \pi \omega_n^2} |F_\pi(\omega_n)|^2\left[ k_n \delta_{11}^\prime(k_n) + 
                               \frac{k_nL}{2\pi} \phi^\prime\left( \frac{k_nL}{2\pi} \right) \right]^{-1} ~ .
     \label{eq:An}
 \ee
For our purposes all we need is a parametrization of the timelike pion form factor $F_\pi(\omega) = |F_\pi(\omega_n)| e^{i\delta_{11}}$, where its phase coincides with the scattering phase shift according to the Watson theorem.
The most popular parametrization is the Gounaris-Sakurai (GS) one~\cite{Gounaris:1968mw}, which is based on the dominance of the $\rho$ resonance in the amplitude of the pion-pion P-wave elastic scattering (with total isospin $1$), namely
\be
     F_\pi^{(GS)}(\omega) = \frac{M_\rho^2 - A_{\pi \pi}(0)}{M_\rho^2 - \omega^2 - A_{\pi \pi}(\omega)} ~ ,
     \label{eq:Fpi_GS}
\ee
where the (twice-subtracted~\cite{Gounaris:1968mw}) pion-pion amplitude $A_{\pi \pi}(\omega)$ is given by
\be
    A_{\pi \pi}(\omega) = h(M_\rho) + (\omega^2 - M_\rho^2) \frac{h^\prime(M_\rho)}{2 M_\rho} - h(\omega) + i \omega \Gamma_{\rho \pi \pi}(\omega)
    \label{eq:Apipi}
\ee
with
\bea
    \label{eq:Gamma_rhopipi}
    \Gamma_{\rho \pi \pi}(\omega) & = & \frac{g_{\rho \pi \pi}^2}{6 \pi} \frac{k^3}{\omega^2} ~ , \\
    \label{eq:homega}
    h(\omega) & = & \frac{g_{\rho \pi \pi}^2}{6 \pi} \frac{k^3}{\omega} \frac{2}{\pi}\mbox{log}\left( \frac{\omega + 2k}{2M_\pi} \right) ~ , \\
    \label{eq:hpomega}
    h^\prime(\omega) & = & \frac{g_{\rho \pi \pi}^2}{6 \pi} \frac{k}{\pi \omega} \left\{ 1 + \left(1 + \frac{2 M_\pi^2}{\omega^2} \right) \frac{\omega}{k} 
                                           \mbox{log}\left(\frac{\omega + 2k}{2 M_\pi} \right) \right\} ~ , \\
    \label{eq:Apipi0}
     A_{\pi \pi}(0) & = & h(M_\rho) - \frac{M_\rho}{2} h^\prime(M_\rho) + \frac{g_{\rho \pi \pi}^2}{6 \pi} \frac{M_\pi^2}{\pi}                                   
\eea
and $k \equiv \sqrt{\omega^2 / 4 - M_\pi^2}$.
By analytic continuation the GS form factor at $\omega = 0$ is normalized to unity, i.e.~$F_\pi^{(GS)}(\omega = 0) = 1$.
The scattering phase shift $\delta_{11}(k)$, i.e.~the phase of the form factor, is given by
\be
    \mbox{cot}\delta_{11}(k) = \frac{M_\rho^2 - \omega^2 - h(M_\rho) - (\omega^2 - M_\rho^2) h^\prime(M_\rho) / (2 M_\rho) +
                                              h(\omega)}{\omega \Gamma_{\rho \pi \pi}(\omega)} ~ .
    \label{eq:delta11}
\ee

\begin{figure}[htb!]
\centering{\scalebox{0.75}{\includegraphics{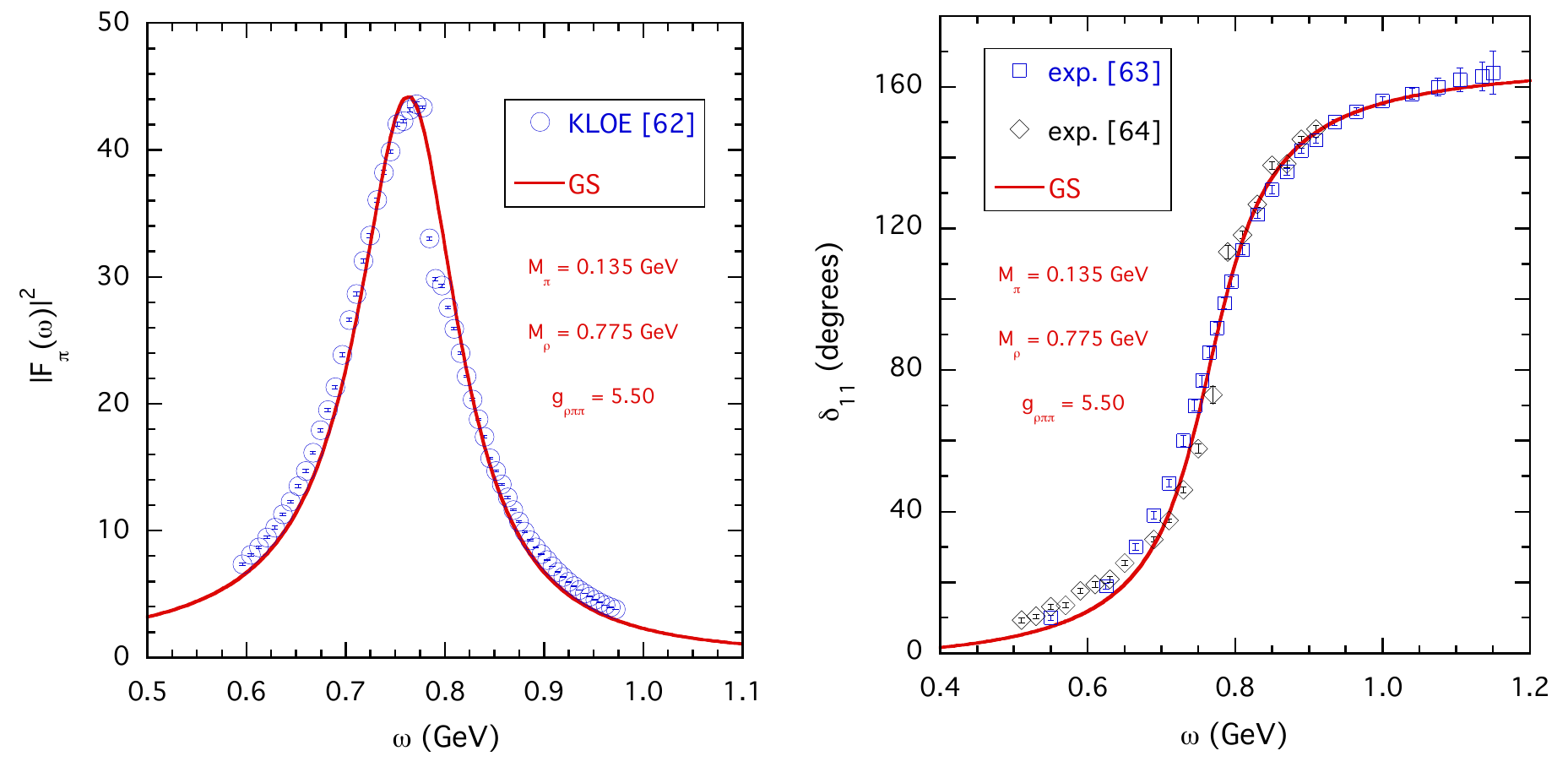}}}
\vspace{-0.25cm}
\caption{\it \small Left panel: the squared timelike pion form factor $|F_\pi(\omega)|^2$ determined by the KLOE experiment~\cite{Babusci:2012rp} from the process $e^+ e^- \to \pi^+ \pi^-$ (dots). Right panel: the experimental values of the scattering phase shift $\delta_{11}$ obtained in Ref.~\cite{Protopopescu:1973sh} (squares) and in Ref.~\cite{Estabrooks:1974vu} (diamonds). The solid lines represent the results of the GS parametrization (\ref{eq:Fpi_GS}-\ref{eq:Gamma_rhopipi}) corresponding to $M_\pi = 0.135$ GeV, $M_\rho = 0.775$ GeV and $g_{\rho \pi \pi} = 5.50$.\hspace*{\fill}}
\label{fig:FPi_GS}
\end{figure}

The GS parametrization contains two parameters: the resonance mass, $M_\rho$, and its strong coupling with two pions, $g_{\rho \pi \pi}$.
At the physical pion point the GS parametrization of the pion form factor provides a reasonable description of the experimental data on the process $e^+ e^- \to \pi^+ \pi^-$, as shown in Fig.~\ref{fig:FPi_GS}.

In what follows we adopt the GS parametrization and treat both $M_\rho$ and $g_{\rho \pi \pi}$ as free parameters to be determined by fitting the vector correlator $V^{(ud)}(t)$.
Note that the GS form factor does not contain any effect of the $\rho - \omega$ mixing.
This is appropriate for our isosymmetric ($m_u = m_d$) QCD lattice setup.

\begin{figure}[htb!]
\centering{\scalebox{0.70}{\includegraphics{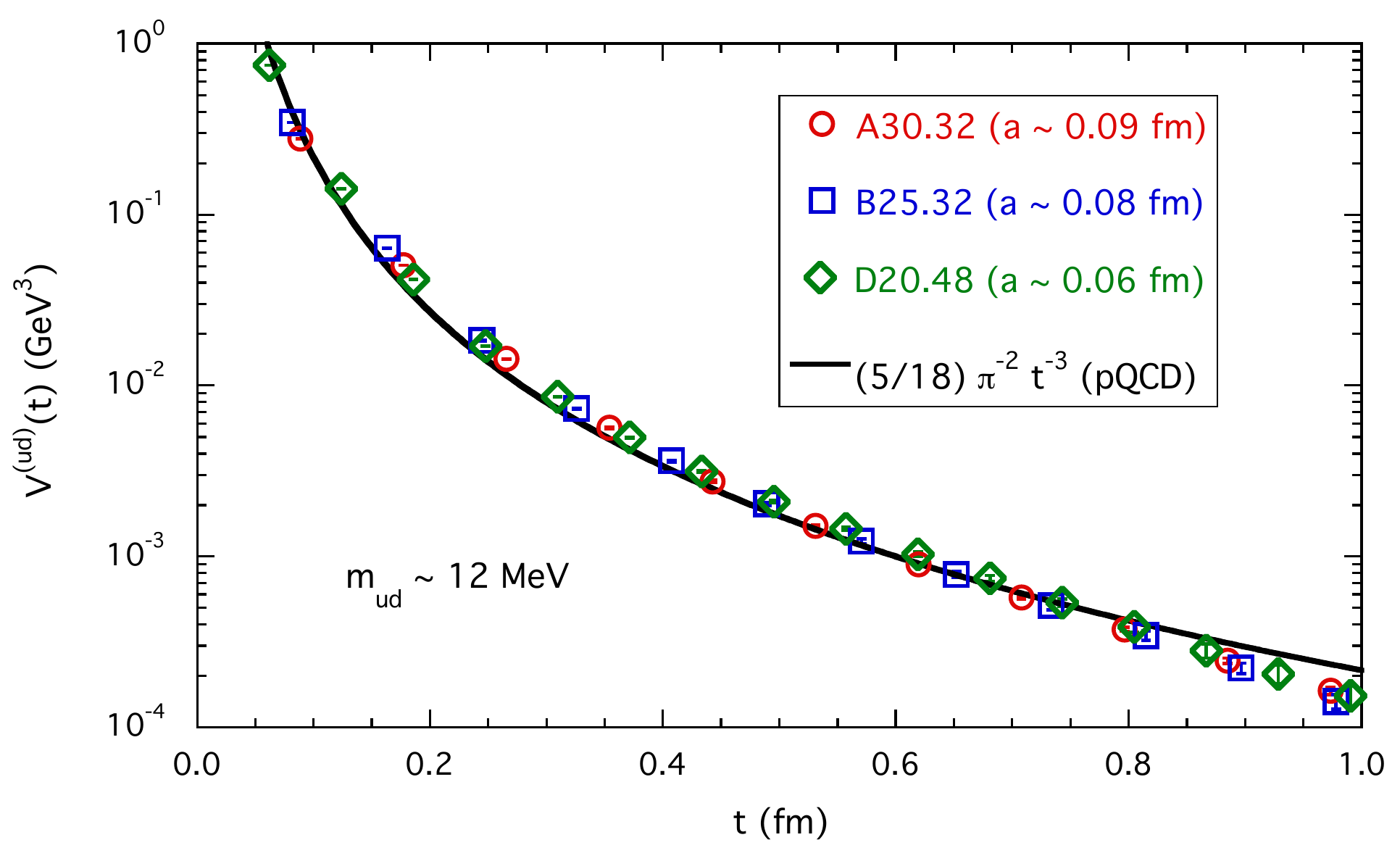}}}
\vspace{-0.25cm}
\caption{\it \small The vector correlator $V^{(ud)}(t)$ in physical units corresponding to the ETMC gauge ensembles specified in the inset, which share an approximate common value of the (renormalized) light-quark mass $m_{ud} \simeq 12$ MeV and differ in the values of the lattice spacing $a$. The solid line represents the pQCD prediction in the massless limit (cf.~Eq.~(3.22) of Ref.~\cite{Giusti:2017jof}).\hspace*{\fill}}
\label{fig:VpQCD}
\end{figure}

We expect that the low-lying states close to the resonance mass can be properly described by the isovector two-pion contribution (\ref{eq:V2pi}).
This means that we may be able to reproduce the vector correlator $V^{(ud)}(t)$ at large time distances.
However, we want to achieve an analytic representation of the vector correlator valid also at low and intermediate time distances.
To this end we resort to an observation made in Ref.~\cite{Giusti:2017jof}, concerning the onset of quark-hadron duality~\cite{SVZ}.
The matching between perturbative QCD (pQCD) and the vector correlator is expected to occur at enough small values of $t$, i.e.~$t << 1 / \Lambda_{QCD} \approx 1$ fm (with $\Lambda_{QCD} \approx 300$ MeV), which correspond to energy scales $>> \Lambda_{QCD}$.
As shown in Ref.~\cite{Giusti:2017jof}, the matching with pQCD occurs instead up to time distances of $\approx 1$ fm.
Such an agreement holds in the case of the light $u$- and $d$-quarks, which can be treated in the massless limit, as well as in the case of the strange and charm quarks, once the corrections due to the nonvanishing quark masses are included. 
The fact that the matching appears to work up to $t \approx 1$ fm is a nice manifestation of the quark-hadron duality {\it \`a la SVZ}, which states that the sum of the contributions of the excited states is dual to the pQCD behaviour~\cite{SVZ}.
The onset of quark-hadron duality in the vector correlator $V^{(ud)}(t)$, evaluated using our lattice data, is illustrated in Fig.~\ref{fig:VpQCD}.

Thus, inspired by the approach of QCD sum rules we introduce a {\em dual} correlator, $V_{dual}(t)$, defined as
 \bea
     V_{dual}(t) & \equiv & \frac{1}{24 \pi^2} \int_{s_{dual}}^\infty ds \sqrt{s} e^{-\sqrt{s} t} R^{pQCD}(s) \nonumber \\
                       & = & \frac{5}{9} \frac{1}{8 \pi^2} \int_{s_{dual}}^\infty ds \sqrt{s} e^{-\sqrt{s} t} 
                                 \left[ \sqrt{1 - \frac{4m_{ud}^2}{s}} \left( 1 + \frac{2 m_{ud}^2}{s} \right) + {\cal{O}}(\alpha_s) \right] \nonumber \\
                       & = & \frac{5}{9} \frac{s_{dual}^{3/2}}{2 \pi^2} \left[ \frac{1}{x^3} e^{-x} \left( 1 + x + \frac{1}{2} x^2 \right) + 
                                {\cal{O}}\left(\frac{m_{ud}^4}{s_{dual}^2}\right) + {\cal{O}}(\alpha_s) \right] ~ ,
     \label{eq:Vdual_0}
 \eea
where $x \equiv \sqrt{s_{dual}} t$ and $s_{dual}$ is an effective threshold above which the hadronic spectral density is considered to be dual to the pQCD prediction $R^{pQCD}(s)$ related to the (one photon) $e^+ e^-$ annihilation cross section into hadrons.

According to Ref.~\cite{SVZ} the value of $\sqrt{s_{dual}}$ is expected to be above the ground-state mass by an amount of the order of $\Lambda_{QCD}$.
Therefore, we assume that
 \be
      s_{dual} = \left( M_\rho + E_{dual} \right)^2
      \label{eq:sdual}
  \ee
with  $E_{dual}$ being treated as a free parameter to be determined by fitting the vector correlator $V^{(ud)}(t)$. 
Furthermore, we introduce in the r.h.s.~of Eq.~(\ref{eq:Vdual_0}) a multiplicative factor $R_{dual}$ in order to take into account perturbative corrections at order ${\cal{O}}(\alpha_s)$ (and beyond), discretization effects and an (expected) slight dependence on the light-quark mass $m_{ud}$\footnote{A more refined treatment of the perturbative and condensate corrections to $V_{dual}(t)$ is left to future developments.}.
 Thus, our final expression for the {\em dual} correlator $V_{dual}(t)$ is
 \be
      V_{dual}(t) = \frac{5}{18 \pi^2} \frac{R_{dual}}{t^3} e^{- (M_\rho + E_{dual}) t} \left[ 1 + (M_\rho + E_{dual}) t + \frac{1}{2} (M_\rho + E_{dual})^2 t^2 \right]
      \label{eq:Vdual} ~ ,
 \ee
where both $R_{dual}$ and $E_{dual}$ are free parameters to be determined by fitting the vector correlator $V^{(ud)}(t)$, while $M_\rho$ is the same parameter appearing in the two-pion contribution (\ref{eq:V2pi}-\ref{eq:An}) through the GS parametrization of the timelike pion form factor (\ref{eq:Fpi_GS}-\ref{eq:Gamma_rhopipi}).

To sum up, our analytic representation of the vector correlator $V^{(ud)}(t)$ is given by the sum of the dual correlator $V_{dual}(t)$ and the two-pion contribution $V_{\pi \pi}(t)$, viz.
 \be
      V_{dual + \pi \pi}(t) = V_{dual}(t) + V_{\pi \pi}(t) ~ ,
      \label{eq:dual+2pi}
 \ee
which contains four free parameters, $R_{dual}$, $E_{dual}$, $M_\rho$ and $g_{\rho \pi \pi}$.
More precisely, we can make use of four dimensionless parameters, namely $R_{dual}$, $E_{dual} / M_\pi$, $M_\rho / M_\pi$ and $g_{\rho \pi \pi}$, which will be determined by fitting the vector correlator $V^{(ud)}(t)$ separately for each of the 17 ETMC gauge ensembles of Table \ref{tab:simudetails}.
In this way the fitting procedure can be carried out entirely in lattice units without requiring the knowledge of the value of the lattice spacing (i.e., the four parameters $R_{dual}$, $E_{dual} / M_\pi$, $M_\rho / M_\pi$ and $g_{\rho \pi \pi}$ are not sensitive to the uncertainty of the scale setting).
We find that the inclusion of the (lowest) four two-pion energy levels $\omega_n$ in Eq.~(\ref{eq:V2pi}) turns out to be sufficient for all of the ETMC ensembles\footnote{We have explicitly checked that using the (lowest) eight energy levels in Eq.~(\ref{eq:V2pi}) yield results for the four parameters $R_{dual}$, $E_{dual}$, $M_\rho$ and $g_{\rho \pi \pi}$, which differ well below the uncertainties.}.

By means of the analytic representation (\ref{eq:dual+2pi}) we reproduce accurately the lattice data for the vector correlator $V^{(ud)}(t)$ for $t \gtrsim 0.2$ fm for all ETMC ensembles.
The fitting region is extended up to larger values of $t$, where the statistical uncertainties of the lattice correlator $V^{(ud)}(t)$ do not exceed $\simeq 10\%$ (i.e., $ t \lesssim 1.7 \div 2.0$ fm).

\begin{figure}[htb!]
\centering{\scalebox{0.70}{\includegraphics{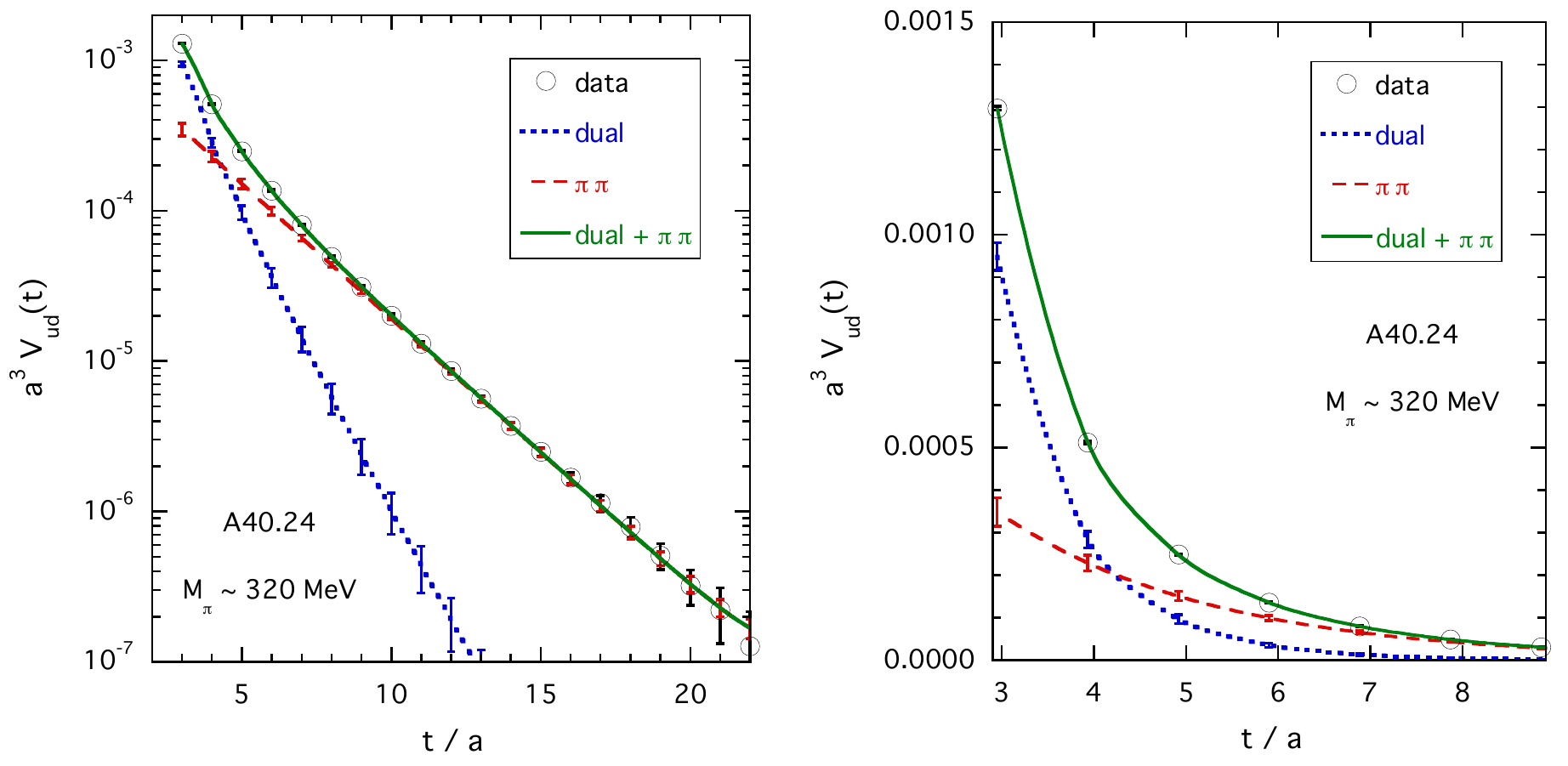}}}
\vspace{-0.25cm}
\caption{\it \small The vector correlator $V^{(ud)}(t)$ (in lattice units) in the case of the gauge ensemble A40.24 corresponding to a pion mass of $\simeq 320$ MeV versus the time distance $t$ (in lattice units). The blue dotted and the red dashed lines represent respectively the contributions of the dual correlator $V_{dual}(t)$ and of the two-pion correlator $V_{\pi \pi}(t)$. The green solid line is the sum of the two contributions. In the left panel a logarithmic scale is used, while in the right panel the region of low values of $t$ is better highlighted using a linear scale. Errors are statistical only.\hspace*{\fill}}
\label{fig:fits_A4024}
\end{figure}

\begin{figure}[htb!]
\centering{\scalebox{0.70}{\includegraphics{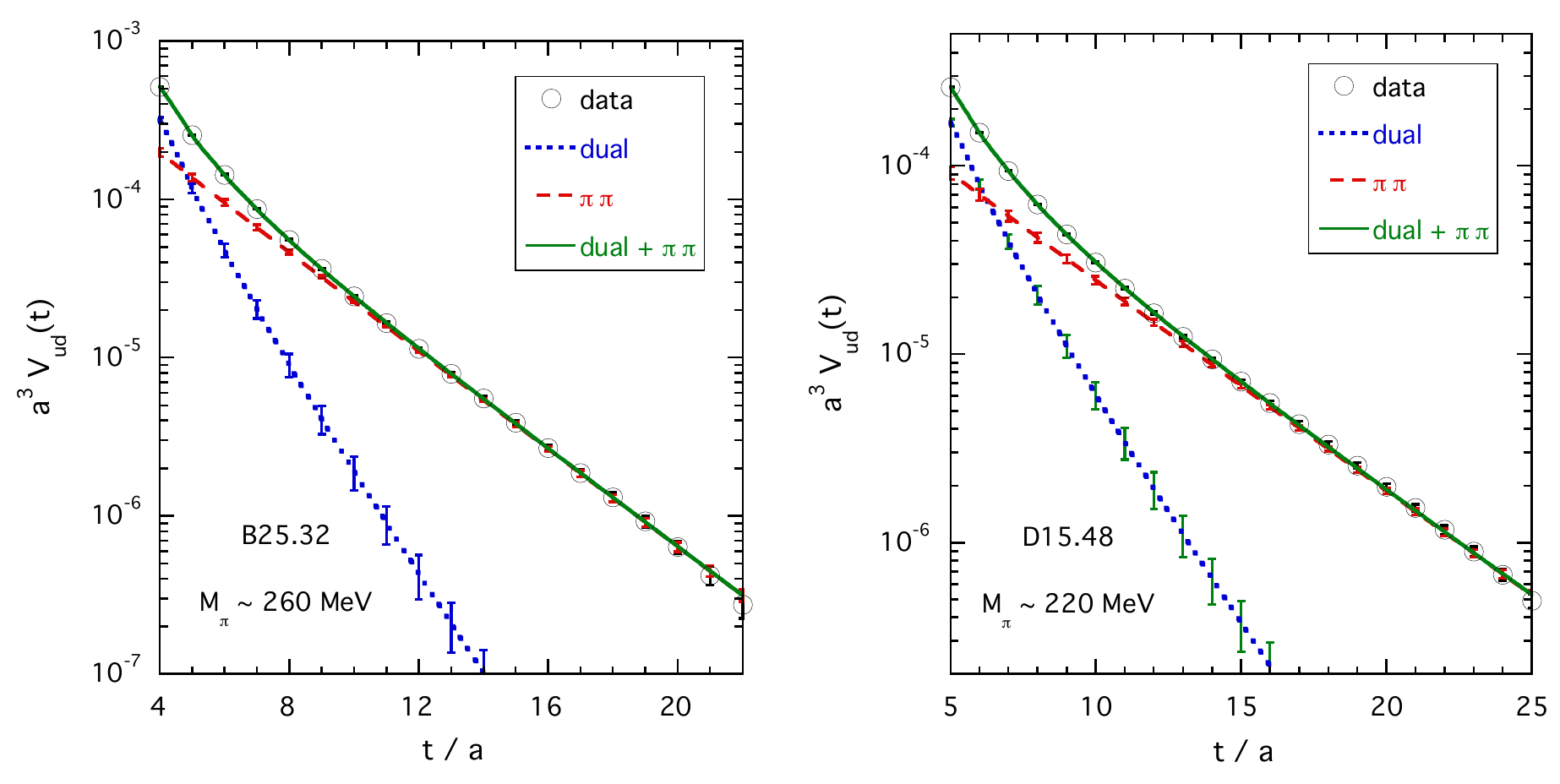}}}
\vspace{-0.25cm}
\caption{\it \small The same as in the left panel of Fig.~\ref{fig:fits_A4024}, but in the case of the gauge ensembles B25.32 and D15.48 corresponding to $M_\pi \simeq 260$ and $\simeq 220$ MeV, respectively. Errors are statistical only.\hspace*{\fill}}
\label{fig:fits_corr}
\end{figure}

\begin{figure}[htb!]
\centering{\scalebox{0.70}{\includegraphics{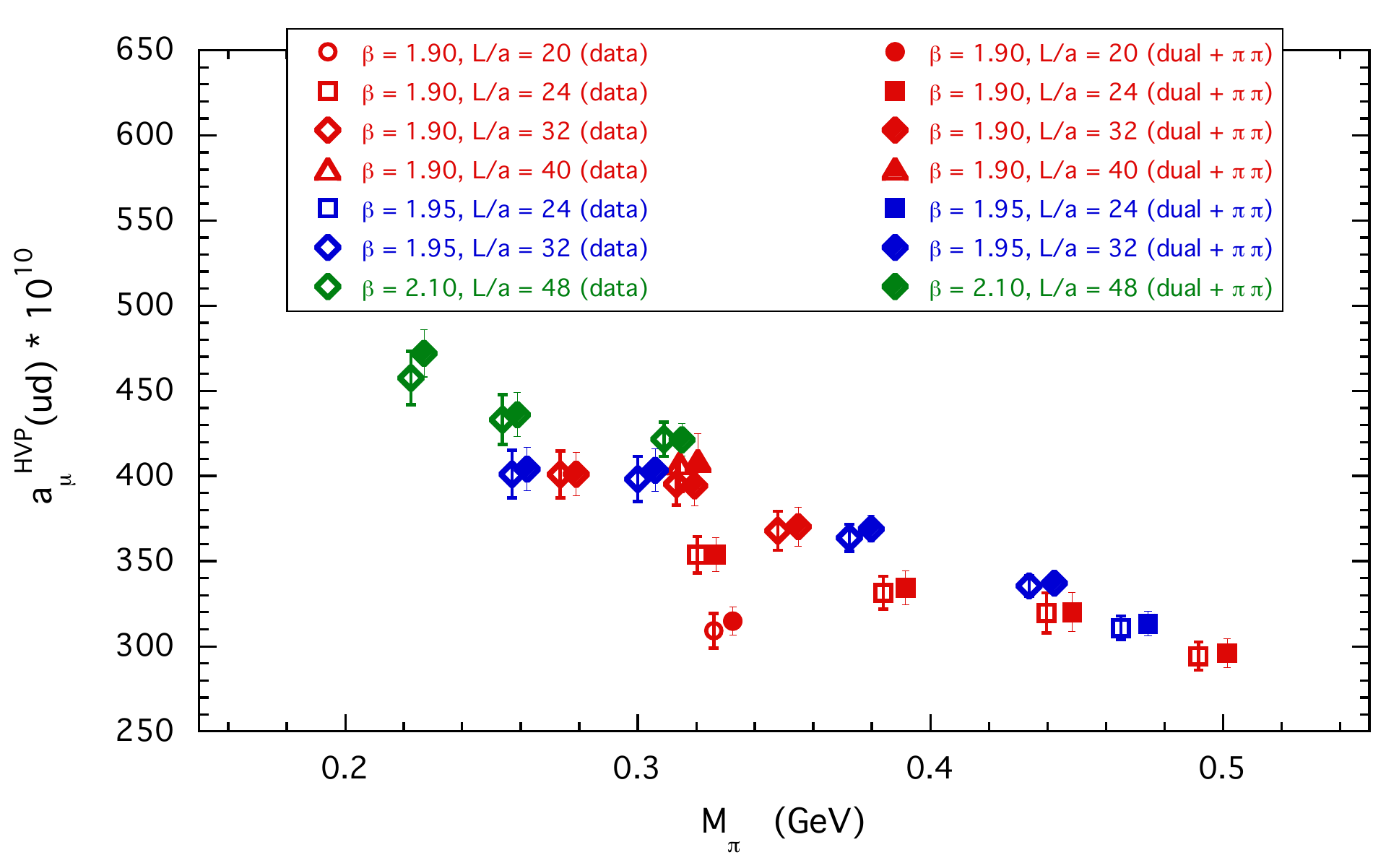}}}
\vspace{-0.25cm}
\caption{\it \small The (connected) light-quark contribution to the muon HVP, $a_\mu^{\rm HVP}(ud)$, evaluated for all the ETMC gauge ensembles of Table \ref{tab:simudetails}. Empty markers correspond to Eq.~(\ref{eq:decomposition}), where the lattice data for the vector correlator $V^{(ud)}(t)$ are directly used. Full markers are the results of Eq.~(\ref{eq:amud_rep}), where the analytic representation (\ref{eq:dual+2pi}) is adopted. For the latter case the points have been shifted horizontally for a better readability.\hspace*{\fill}}
\label{fig:comparison}
\end{figure}

The quality of the fits is illustrated in Figs.~\ref{fig:fits_A4024}-\ref{fig:fits_corr} in the case of few ETMC gauge ensembles and it is nicely confirmed by the comparison, shown in Fig.~\ref{fig:comparison}, between the values of $a_\mu^{\rm HVP}(ud)$, evaluated using Eq.~(\ref{eq:decomposition}), and those corresponding to the analytic representation (\ref{eq:dual+2pi}), namely
\be
     a_\mu^{\rm HVP}(ud)|_{dual+\pi\pi} = 4 \alpha_{em}^2 \sum_{t=0}^\infty f(t) \left[ V_{dual}(t) + V_{\pi \pi}(t) \right]~ .
     \label{eq:amud_rep}
\ee
The high-level accuracy obtained for the reproduction of the vector correlator $V^{(ud)}(t)$ using the analytic representation (\ref{eq:dual+2pi}) guarantees that the calculated values of $a_\mu^{\rm HVP}(ud)$ differs form the lattice data less than one standard deviation. 

We point out that for all the ETMC ensembles of Table~\ref{tab:simudetails} the first noninteracting two-pion energy level, given by $2 \sqrt{M_\pi^2 + (2 \pi / L)^2}$, is always well above the position of the resonance mass $M_\rho$.
Due to the residual strong interaction between the two pions the first energy level $\omega_{n = 1}$ satisfying the L\"uscher condition (\ref{eq:kn}) turns out to be slightly below  $M_\rho$.
This feature justifies the use of a single exponential fit in Eq.~(\ref{eq:decomposition}), at least for the ETMC ensembles of Table~\ref{tab:simudetails}.
Such a situation changes as the simulated pion mass decreases and the single exponential fit is completely ruled out at the physical pion point (see later section \ref{sec:physical}).

Before closing this section, we address the issue of possible correlations of the vector correlator $V^{(ud)}(t)$ at nearby values of $t$.
To this end we have repeated our fitting procedure with reduced numbers of data corresponding to including one out of two (or three) subsequent lattice points. 
The results obtained for the four parameters $R_{dual}$, $E_{dual} / M_\pi$, $M_\rho / M_\pi$ and $g_{\rho \pi \pi}$ differ within approximately one standard deviation, as shown in Table \ref{tab:autocorrelations} in the case of few ETMC gauge ensembles.
\begin{table}[hbt!]
\begin{center}
\renewcommand{\arraystretch}{1.10}
\begin{tabular}{||c||c|c||c|c||c|c|c||}
\hline
parameter & $A40.24$ & $A40.24$ & $B55.32$ & $B55.32$ & $D30.48$ & $D30.48$ & $D30.48$ \\
\cline{2-8}
                  & all & 1 out of 2 & all & 1 out of 2 & all & 1 out of 2 & 1 out of 3 \\
\hline \hline
           $R_{dual}$ & $1.44 ~~ (4)$ & $1.43  ~ ~(4)$ & $1.39  ~ (2)$ & $1.39  ~~ (3)$ & $1.21  ~~ (1)$ & $1.20  ~ (1)$ & $1.18  ~~ (2)$ \\
\hline
$E_{dual} / M_\pi$ & $2.22  ~ (23)$ & $2.13 ~  (27)$ & $1.95  ~ (9)$ & $1.96  ~ (10)$ & $1.84  ~ (10)$ & $1.77  ~ (11)$ & $1.65  ~ (13)$  \\
\hline
   $M_\rho / M_\pi$ & $2.77  ~~ (9)$ & $2.76  ~ (10)$ & $2.44  ~ (2)$ & $2.44  ~~ (2)$ & $2.76  ~~ (4)$ & $2.74  ~ (4)$ & $2.73  ~~ (4)$  \\
\hline
    $g_{\rho \pi \pi}$ & $5.22  ~~ (9)$ & $5.25  ~ (11)$ & $4.98  ~ (2)$ & $4.98  ~~ (3)$ & $5.04  ~~ (4)$ & $5.07  ~ (4)$ & $5.10  ~~ (5)$  \\
\hline   
\end{tabular}
\renewcommand{\arraystretch}{1.0}
\end{center}
\vspace{-0.25cm}
\caption{\it \small Values of the four parameters $R_{dual}$, $E_{dual} / M_\pi$, $M_\rho / M_\pi$ and $g_{\rho \pi \pi}$ obtained by fitting the vector correlator $V^{(ud)}(t)$ by including all subsequent timeslices (all) or one out of two (or three) subsequent lattice points in the case of the gauge ensembles A40.24, B55.32 and D30.48.\hspace*{\fill}}
\label{tab:autocorrelations}
\end{table}

The results for the four parameters $R_{dual}$, $E_{dual} / M_\pi$, $M_\rho / M_\pi$ and $g_{\rho \pi \pi}$ obtained in the case of all the ETMC ensembles will be shown later in Figs.~\ref{fig:dual_ETMC}-\ref{fig:2pion_ETMC}.

\section{Subtraction of FVEs}
\label{sec:FVE}

We start the analysis of FVEs by considering the four ensembles A40.XX, which share the same quark mass ($m_{ud} \simeq 17$ MeV) and lattice spacing ($a \simeq 0.089$ fm) and differ only in the lattice size $L$, namely $XX = 20, 24, 32$ and $40$ (see Table~\ref{tab:simudetails}).

\subsection{Ensembles A40.XX}
\label{sec:FVE_A40XX}

\begin{figure}[htb!]
\centering{\scalebox{0.70}{\includegraphics{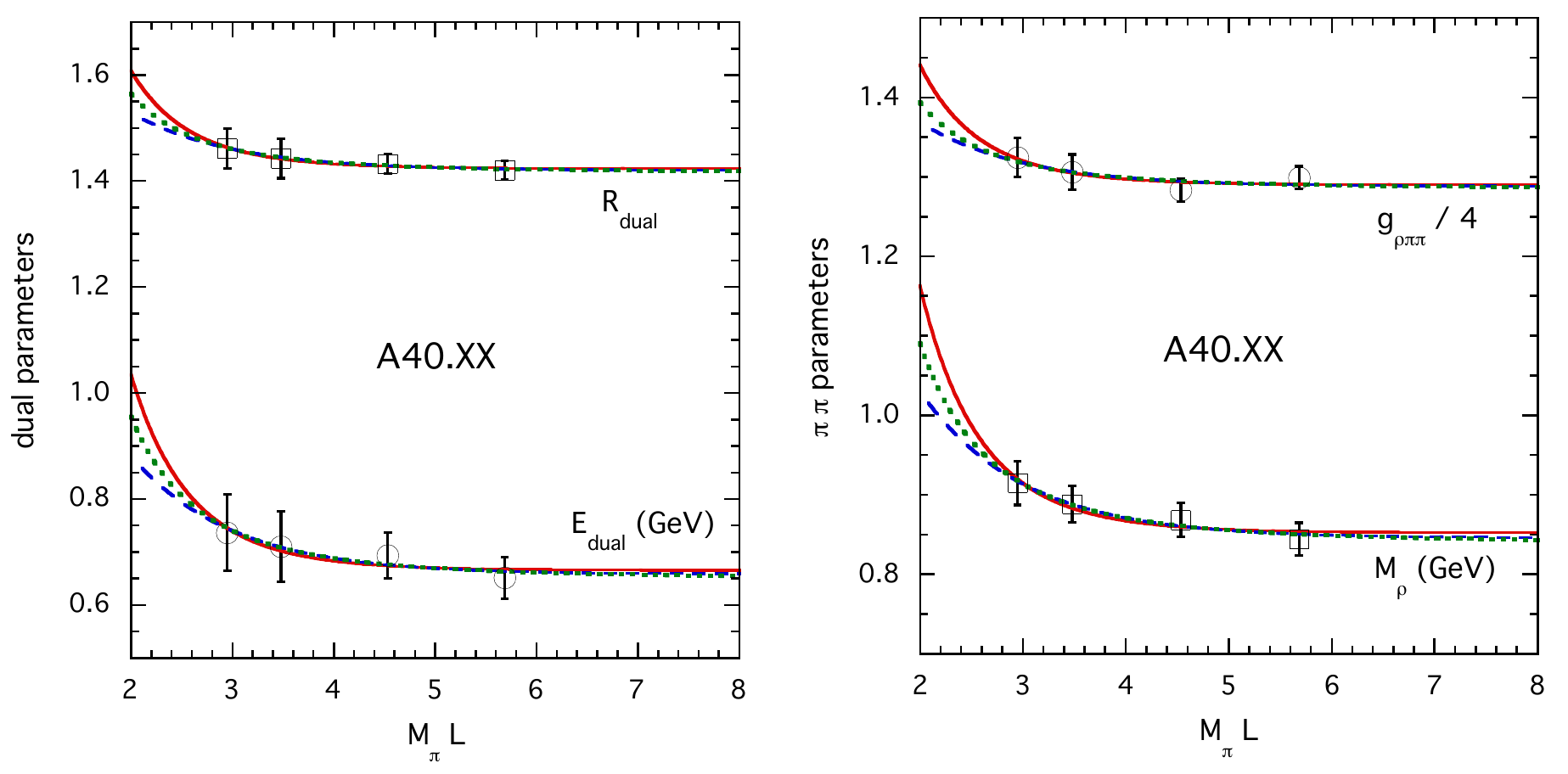}}}
\vspace{-0.25cm}
\caption{\it \small Left panel: the dual parameters $R_{dual}$ and $E_{dual}$ (given in physical units) versus $M_\pi L$, appearing in Eq.~(\ref{eq:Vdual}), for the four ensembles A40.XX ($m_{ud} \simeq 17$ MeV and $a \simeq 0.089$ fm). Right panel: the parameters $M_\rho$ (given in physical units) and $g_{\rho \pi \pi} / 4$ versus $M_\pi L$, appearing in the two-pion contribution (\ref{eq:V2pi}). The solid and dashed lines correspond to the exponentially-suppressed Ansatz (\ref{eq:expon}) with $\alpha = 3/2$ and $\alpha = 0$, respectively.  The dotted lines correspond to the power-suppressed Ansatz (\ref{eq:power}).\hspace*{\fill}}
\label{fig:parms_A40XX}
\end{figure}

The values of the four parameters $R_{dual}$, $E_{dual}$, $M_\rho$ and $g_{\rho \pi \pi}$ obtained by fitting the vector correlator $V^{(ud)}(t)$ are shown in Fig.~\ref{fig:parms_A40XX}.
It can be seen that the FVEs on all the fitting parameters are definitely more limited with respect to those observed for $a_\mu^{\rm HVP}(ud)$ in the right panel of Fig.~\ref{fig:amul_ETMC}.
This fact allows for a good control of the values of the four parameters in the infinite volume limit, as shown in Fig.~\ref{fig:parms_A40XX} by the solid, dashed and dotted lines, whose differences are well within the uncertainties.
The solid and dashed lines correspond to the exponentially-suppressed Ansatz
 \be
      P = P^\infty \left[ 1 + F_P \frac{e^{-M_\pi L}}{(M_\pi L)^\alpha} \right]
      \label{eq:expon}
 \ee
with $\alpha = 3/2$ and $\alpha = 0$, respectively.
In Eq.~(\ref{eq:expon}) $P$ stands for $\{ R_{dual},  E_{dual}, M_\rho, g_{\rho \pi \pi} \}$, while $P^\infty$ and $F_P$ are fitting parameters.
The dotted lines correspond instead to the power-suppressed Ansatz
 \be
      P = P^{\prime \infty} \left[ 1 + \frac{F_P^\prime}{(M_\pi L)^3} \right] ~ .
      \label{eq:power}
 \ee
Besides the four parameters $R_{dual}$, $E_{dual}$, $M_\rho$ and $g_{\rho \pi \pi}$, also the simulated pion mass $M_\pi$ suffers from FVEs, which have been thoroughly investigated in Ref.~\cite{Carrasco:2014cwa} using the resummed ChPT approach of Refs.~\cite{Colangelo:2005gd,Colangelo:2010cu}.
For the purposes of the present work it suffices to consider for $M_\pi^2$ the exponentially-suppressed Anzatz (\ref{eq:expon}) with $\alpha = 3/2$, as suggested by the asymptotic behavior of NLO ChPT in the $p$-regime.

Once the infinite volume limits $R_{dual}^\infty$, $E_{dual}^\infty$, $M_\rho^\infty$, $g_{\rho \pi \pi}^\infty$ and $M_\pi^\infty$ have been determined, we need to specify the infinite-volume limit of our ``dual+$\pi\pi$" representation
 \be
      V_{dual + \pi \pi}^\infty(t) = V_{dual}^\infty(t) + V_{\pi \pi}^\infty(t) ~ .
      \label{eq:dual+2pi_vol}
 \ee
For the dual contribution one has straightforwardly
 \be
      V_{dual}^\infty(t) = \frac{5}{18 \pi^2} \frac{R_{dual}^\infty}{t^3} e^{- (M_\rho^\infty + E_{dual}^\infty) t} \left[ 1 + (M_\rho^\infty + E_{dual}^\infty) t +
                                    \frac{1}{2} (M_\rho^\infty + E_{dual}^\infty)^2 t^2 \right] ~ ,
      \label{eq:Vdual_vol} 
 \ee
while the two-pion contribution in the infinite-volume limit becomes~\cite{Meyer:2011um}
 \be
     V_{\pi \pi}(t)_ { ~ \overrightarrow{L \to \infty} ~ } V_{\pi \pi}^\infty(t) = \frac{1}{48 \pi^2} \int_{2M_\pi^\infty}^\infty d\omega ~ \omega^2 
                         \left[ 1 - \frac{(2 M_\pi^\infty)^2}{\omega^2} \right]^{3/2} |F_\pi^\infty(\omega)|^2 e^{-\omega t} ~ ,
     \label{eq:V2pi_vol}
 \ee
where $|F_\pi^\infty(\omega)|$ can be calculated from the GS parametrization (\ref{eq:Fpi_GS}-\ref{eq:Gamma_rhopipi}) using $M_\rho^\infty$, $g_{\rho \pi \pi}^\infty$ and $M_\pi^\infty$.

\begin{figure}[htb!]
\centering{\scalebox{0.75}{\includegraphics{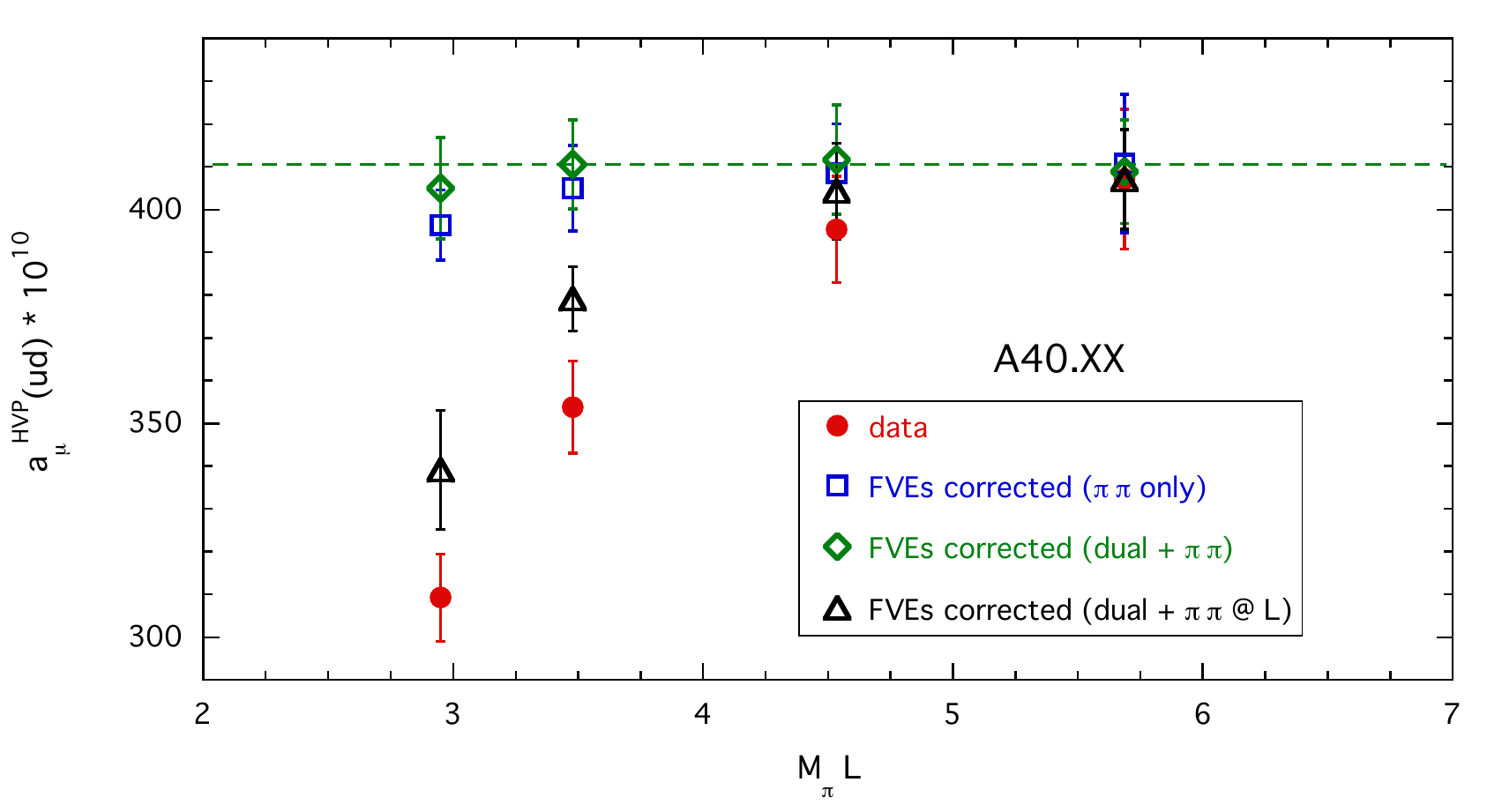}}}
\vspace{-0.25cm}
\caption{\it \small Lattice data in the case of the four ensembles A40.XX (red dots) versus $M_\pi L$. The blue squares and the green diamonds correspond respectively to the data corrected by FVEs according to Eqs.~(\ref{eq:amud_noFVE}-\ref{eq:deltaFVE}), evaluated by including either the $\pi \pi$ contribution only or the ``dual + $\pi \pi$'' terms and by using in Eqs.~(\ref{eq:Vdual_vol}-\ref{eq:V2pi_vol}) the infinite-volume values $R_{dual}^\infty$, $E_{dual}^\infty$, $M_\rho^\infty$, $g_{\rho \pi \pi}^\infty$ and $M_\pi^\infty$. The dashed line is a constant fit to the green points. The black triangles represent the data corrected by the FVEs evaluated using in Eqs.~(\ref{eq:Vdual_vol}-\ref{eq:V2pi_vol}) the values of $R_{dual}$, $E_{dual}$, $M_\rho$, $g_{\rho \pi \pi}$ and $M_\pi$ obtained at each lattice size $L$ (see text).\hspace*{\fill}}
\label{fig:amud_A40XX}
\end{figure}

We can now correct the lattice data for $a_\mu^{\rm HVP}(ud)$, obtained at finite volume by means of Eq.~(\ref{eq:decomposition}), for the FVEs evaluated using our representation of the vector correlator $V^{(ud)}(t)$ at infinite volume, Eqs.~(\ref{eq:Vdual_vol}-\ref{eq:V2pi_vol}), and the one at finite volume, Eqs.~(\ref{eq:V2pi}) and (\ref{eq:Vdual}), namely 
 \bea
      \label{eq:amud_noFVE}
      a_\mu^{\rm HVP}(ud)|_{L \to \infty} & = & a_\mu^{\rm HVP}(ud) + \Delta_{FVE} ~ a_\mu^{\rm HVP}(ud) ~ , \\[2mm]
      \label{eq:deltaFVE}
      \Delta_{FVE} a_\mu^{\rm HVP}(ud) & = & 4 \alpha_{em}^2 \sum_{t=0}^\infty f(t) \left[ V_{dual}^\infty(t) - V_{dual}(t) + V_{\pi \pi}^\infty(t) - V_{\pi \pi}(t)\right] ~ .
 \eea
The results obtained in the case of the ensembles A40.XX are shown in Fig.~\ref{fig:amud_A40XX}.
We observe that most of the FVE correction comes from the $\pi \pi$ contribution. 
The small residual FVEs can be almost totally taken into account by adding the FVEs related to the dual contribution.
We point out that in order to remove properly the FVEs it is important to use in Eqs.~(\ref{eq:Vdual_vol}-\ref{eq:V2pi_vol}) the infinite-volume values $R_{dual}^\infty$, $E_{dual}^\infty$, $M_\rho^\infty$, $g_{\rho \pi \pi}^\infty$ and $M_\pi^\infty$.
Indeed, if one uses instead the finite volume values (as done, e.g., in Ref.~\cite{DellaMorte:2017dyu}), the correction (\ref{eq:deltaFVE}) may be largely underestimated, as shown by the black triangles in Fig.~\ref{fig:amud_A40XX}.

We have explicitly checked the dependence of our FVE correction (\ref{eq:amud_noFVE}) on the parametrization adopted for the timelike pion form factor $F_\pi(\omega)$.
To this end we keep the $\rho$-meson dominance and consider two simple Breit-Wigner forms in which either $\Gamma_{\rho \pi \pi} = const.$ (labeled hereafter as BW) or $\Gamma_{\rho \pi \pi} \propto k$ (labeled as BW$^\prime$) instead of the GS width (\ref{eq:Gamma_rhopipi}).
Correspondingly, the real part of the two-pion amplitude $A_{\pi \pi}(\omega)$ has been calculated using twice-subtracted dispersion relations, as in the case of the GS parametrization.
We have considered also the approximation of neglecting the real part of $A_{\pi \pi}(\omega)$.
The fitting procedure of the vector correlator $V^{(ud)}(t)$ corresponding to the four ensembles A40.XX has been repeated for all the parametrizations of the pion form factor and the values obtained for the dual and $\pi \pi$ parameters have been extrapolated to the infinite volume limit.
The results for $a_\mu^{\rm HVP}(ud)|_{L \to \infty}$, corresponding to Eqs.~(\ref{eq:amud_noFVE}-\ref{eq:deltaFVE}), are collected in Table \ref{tab:Fpi_parm}.

\begin{table}[hbt!]
\begin{center}
\renewcommand{\arraystretch}{1.10}
\begin{tabular}{||c||c|c|c|c||c||}
\hline
parametrization of $F_\pi(\omega)$ & ~ A40.20 ~ & ~ A40.24 ~ & ~ A40.32~  & ~ A40.40 ~ & dual + $\pi \pi$ ($L \to \infty$) \\
\hline \hline
GS with $\mbox{Re}{A_{\pi \pi}} \neq 0$ & 405 (12) & 411 (10) & 412 (13) & 409 (12) & 411 (13) \\
\hline
GS with $\mbox{Re}{A_{\pi \pi}} = 0$ & 406 (13) & 411 (11) & 413 (13) & 410 (13) & 411 (13) \\
\hline \hline
BW with $\mbox{Re}{A_{\pi \pi}} \neq 0$ & 404 (12) & 410 (10) & 411 (13) & 408 (12) & 410 (12) \\
\hline
BW with $\mbox{Re}{A_{\pi \pi}} = 0$ & 404 (12) & 409 (10) & 410 (13) & 407 (12) & 409 (12) \\
\hline \hline
BW$^\prime$ with $\mbox{Re}{A_{\pi \pi}} \neq 0$ & 405 (12) & 410 (10) & 412 (13) & 409 (12) & 410 (12) \\
\hline
BW$^\prime$ with $\mbox{Re}{A_{\pi \pi}} = 0$ & 404 (12) & 411 (10) & 411 (12) & 408 (13) & 410 (13) \\
\hline \hline
\end{tabular}
\renewcommand{\arraystretch}{1.0}
\end{center}
\vspace{-0.25cm}
\caption{\it \small Values of $a_\mu^{\rm HVP}(ud)|_{L \to \infty}$ obtained using Eqs.~(\ref{eq:amud_noFVE}-\ref{eq:deltaFVE}) for the four ensembles A40.XX, adopting different parametrizations of the timelike pion form factor $F_\pi(\omega)$. Besides the GS one [see Eq.~(\ref{eq:Fpi_GS})], two simple Breit-Wigner forms in which either $\Gamma_{\rho \pi \pi} = const.$ {\rm (BW)} or  $\Gamma_{\rho \pi \pi} \propto k$ {\rm (BW$^\prime$)} have been considered; moreover, the real part of the corresponding (twice-subtracted) two-pion amplitude $A_{\pi \pi}(\omega)$ is either included or excluded. The last column represents the values of $a_\mu^{\rm HVP}(ud)|_{dual+\pi\pi}$ obtained in the infinite volume limit for each parametrization of $F_\pi(\omega)$.\hspace*{\fill}}
\label{tab:Fpi_parm}
\end{table}
The changes due to different parametrizations of the timelike pion form factor $F_\pi(\omega)$ are quite small and they do not exceed $\simeq 0.5 \%$.
This finding may be due to the fact that in calculating $\Delta_{FVE} a_\mu^{\rm HVP}(ud)$ (see Eq.~(\ref{eq:deltaFVE})) the differences [$V_{\pi \pi}^\infty(t) - V_{\pi \pi}(t)$] are expected to be less sensitive to the specific parametrization of the pion form factor than the separate terms.

\begin{figure}[htb!]
\centering{\scalebox{0.75}{\includegraphics{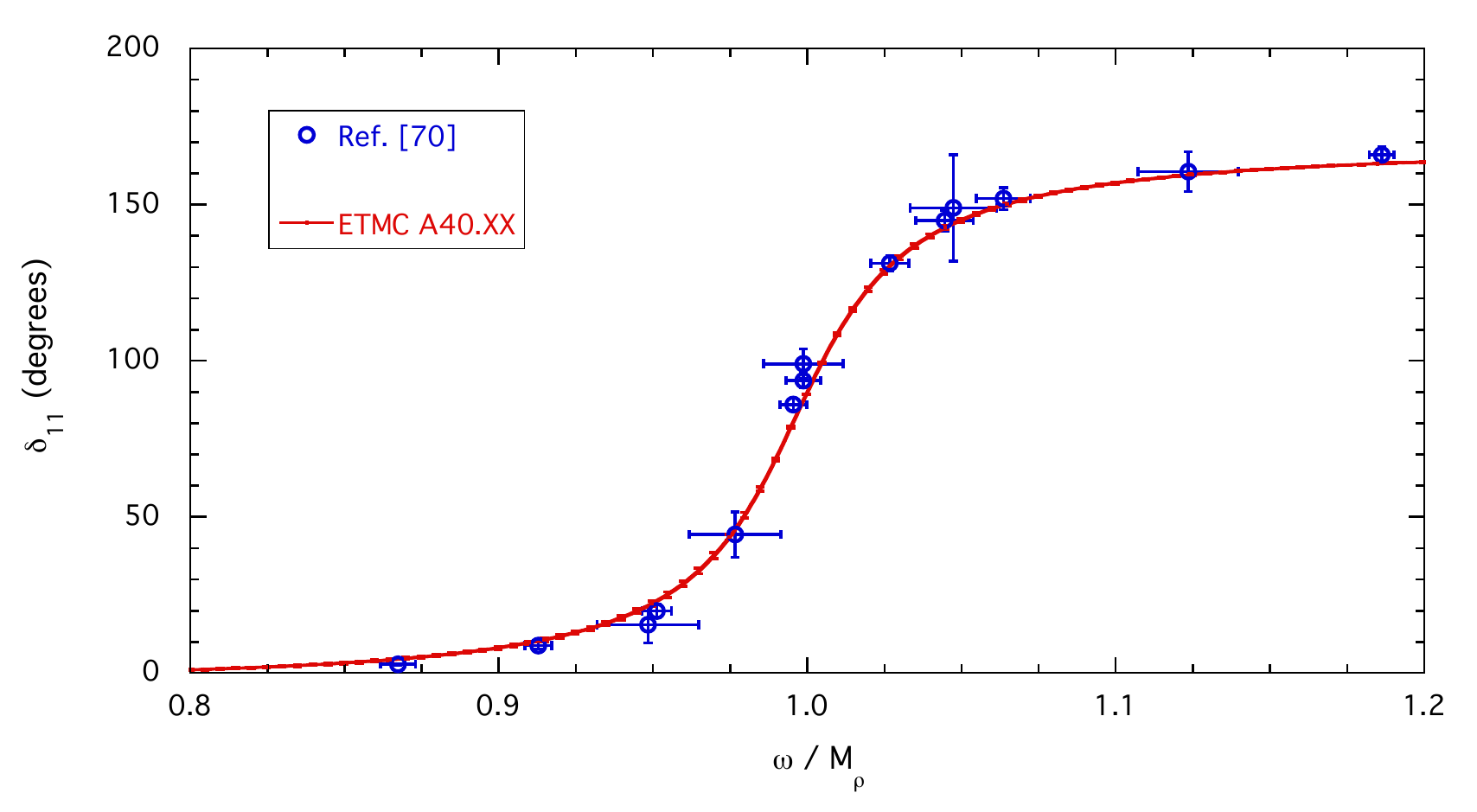}}}
\vspace{-0.25cm}
\caption{\it \small Elastic P-wave $\pi \pi$ scattering phase shift $\delta_{11}$ obtained in Ref.~\cite{Alexandrou:2017mpi} (blue circles) and with our A40.XX ensembles (red curve) versus the dimensionless variable $\omega / M_\rho$. The lattice setup of Ref.~\cite{Alexandrou:2017mpi} corresponds to $N_f = 2+1$ clover fermions with $M_\pi \simeq 320$ MeV, $M_\rho \simeq 800$ MeV, $a \simeq 0.114$ fm and $L \simeq 3.65$ fm. Our A40.XX setup corresponds to $N_f = 2+1+1$ twisted-mass fermions in the infinite volume limit with $M_\pi^\infty \simeq 315 $ MeV, $M_\rho^\infty \simeq 850$ MeV and $a \simeq 0.089$ fm.\hspace*{\fill}}
\label{fig:delta11_A40XX}
\end{figure}

Before closing this subsection, we compare our findings with the results of Ref.~\cite{Alexandrou:2017mpi}, where the elastic P-wave $\pi \pi$ phase shifts $\delta_{11}$ have been extracted from lattice QCD simulations with $N_f = 2 + 1$ flavors of clover fermions.
There the simulated pion mass was $M_\pi \simeq 320$ MeV, which is quite close to the pion mass corresponding to our A40.XX ensembles ($M_\pi^\infty \simeq 315$ MeV).
The phase shifts $\delta_{11}$ found in Ref.~\cite{Alexandrou:2017mpi} are compared in Fig.~\ref{fig:delta11_A40XX} with our A40.XX results corresponding to the infinite volume limit.
The comparison is made in terms of the dimensionless variable $\omega / M_\rho$, which helps in absorbing the different values of the $\rho$-meson mass found in Ref.~\cite{Alexandrou:2017mpi}, $M_\rho \simeq 800$ MeV, and with our A40.XX ensembles, $M_\rho^\infty \simeq 850$ MeV, as well as in absorbing the statistical fluctuations of the $\rho$-meson mass. 
It should be kept in mind that discretization effects are expected to be different between the lattice setup of Ref.~\cite{Alexandrou:2017mpi} and our A40.XX ensembles. Nevertheless, the overall agreement shown in Fig.~\ref{fig:delta11_A40XX} is quite reassuring.

\subsection{ETMC ensembles}
\label{sec:FVE_ETMC}

\begin{figure}[htb!]
\centering{\scalebox{0.75}{\includegraphics{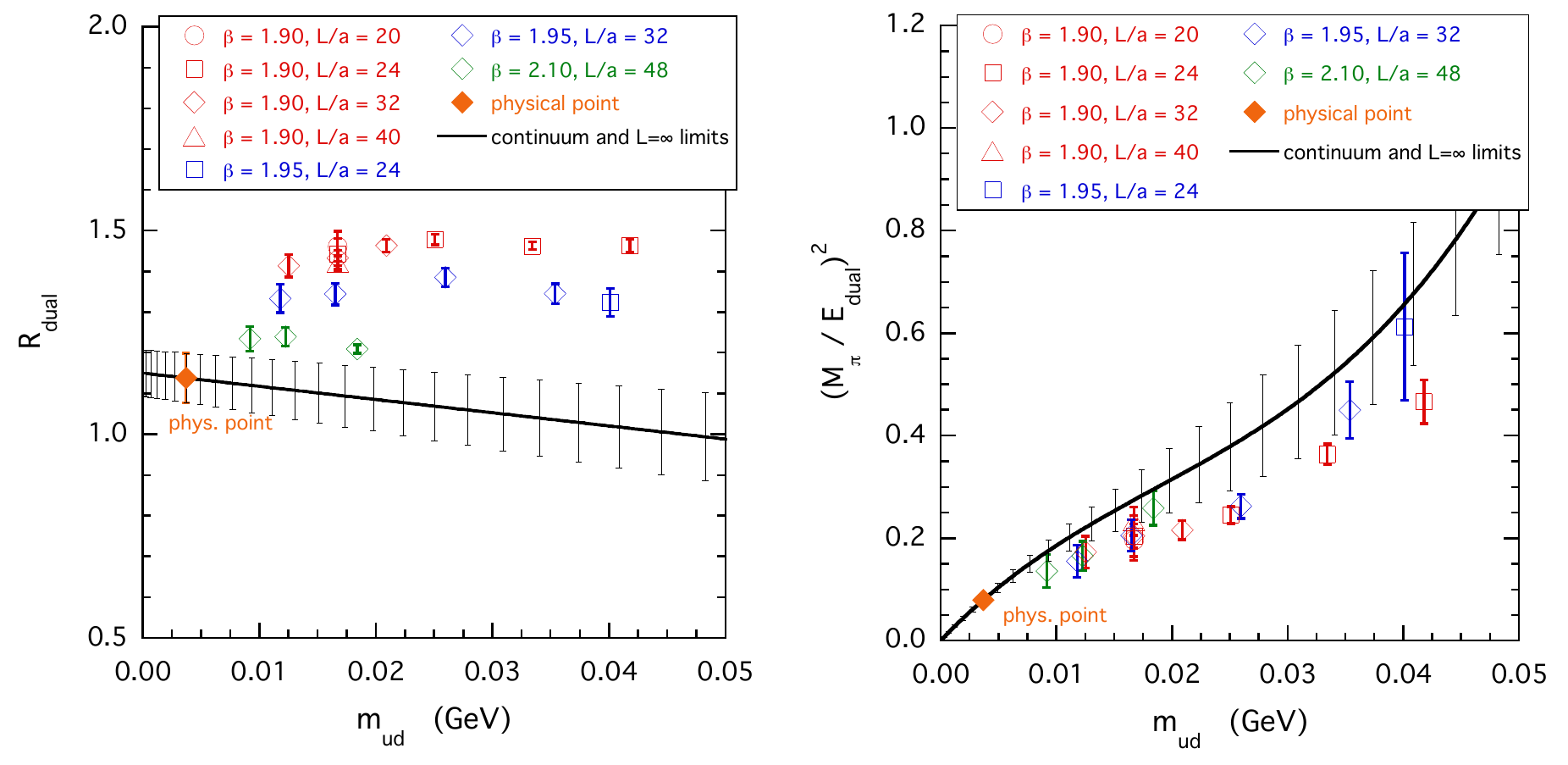}}}
\vspace{-0.25cm}
\caption{\it \small Left panel: the dual parameter $R_{dual}$ versus the renormalized light-quark mass $m_{ud}$ (in the $\overline{MS}(2~GeV)$ scheme) obtained for all the ETMC ensembles of Table \ref{tab:simudetails}. Right panel: the same as in the left panel, but for the dual parameter $(M_\pi / E_{dual})^2$. The solid lines represent respectively the fitting functions (\ref{eq:Rdual_fit}) and (\ref{eq:Edual_fit}) evaluated in the continuum and infinite volume limits. The full (orange) diamonds identify the values of the parameters at the physical pion point, namely $R_{dual}(m_{ud}^{phys}, 0, \infty) = 1.14 ~ (6)$ and $E_{dual}(m_{ud}^{phys}, 0, \infty) = 479 ~ (22)$ MeV.\hspace*{\fill}}
\label{fig:dual_ETMC}
\end{figure}

\begin{figure}[htb!]
\centering{\scalebox{0.75}{\includegraphics{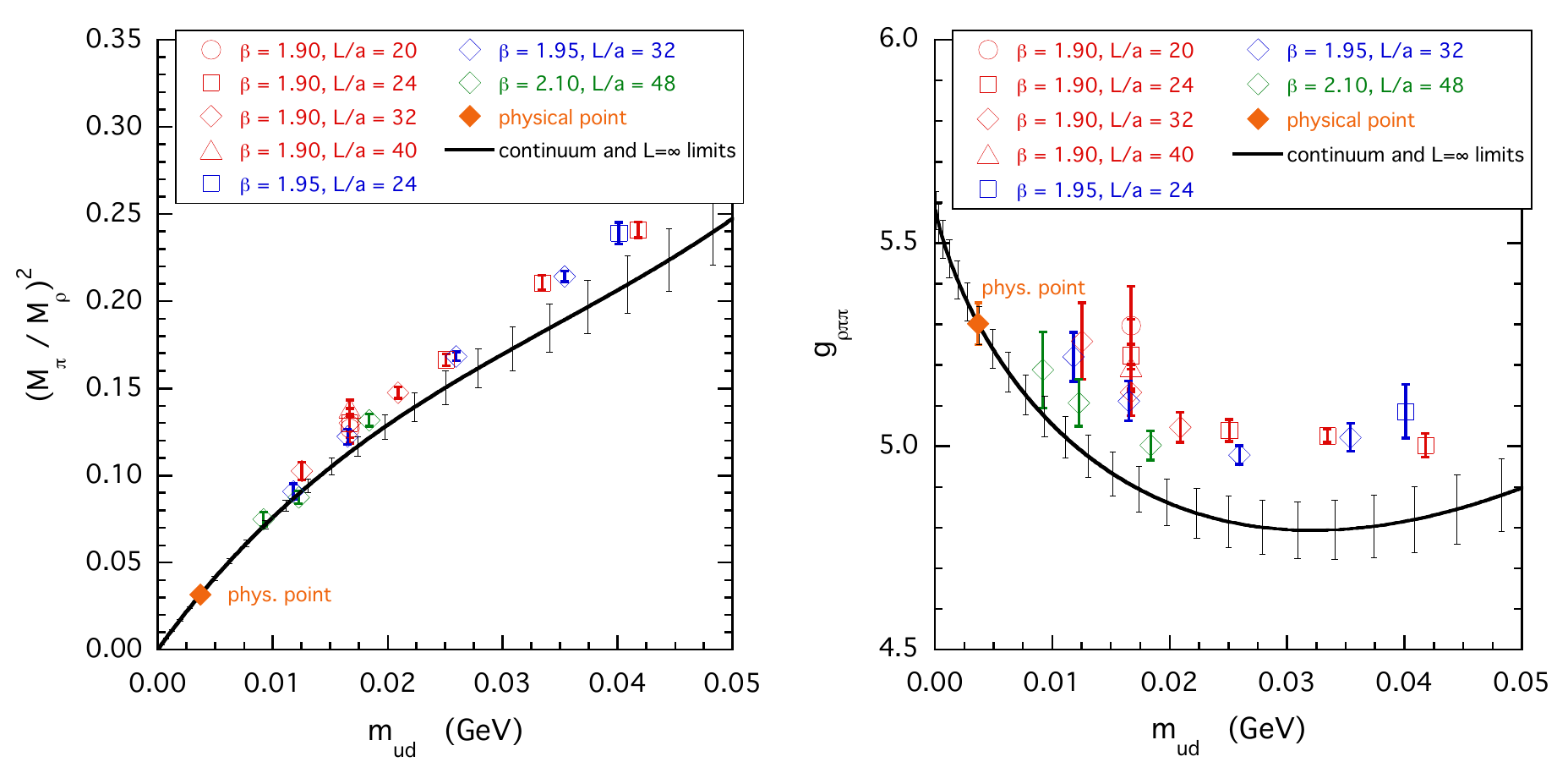}}}
\vspace{-0.25cm}
\caption{\it \small The same as in Fig.~\ref{fig:dual_ETMC}, but for the two-pion parameters $(M_\pi / M_\rho)^2$ and $g_{\rho \pi \pi}$. The solid lines represent respectively the fitting functions (\ref{eq:MV_fit}) and (\ref{eq:gVPP_fit}) evaluated in the continuum and infinite volume limits, namely $M_\rho(m_{ud}^{phys}, 0, \infty) = 760 ~ (19)$ MeV and $g_{\rho \pi \pi}(m_{ud}^{phys}, 0, \infty) = 5.30 ~ (5)$.\hspace*{\fill}}
\label{fig:2pion_ETMC}
\end{figure}

We now address the subtraction of FVEs from the HVP term $a_\mu^{\rm HVP}(ud)$ corresponding to the ETMC ensembles of Table \ref{tab:simudetails}.
The fitting procedure of the vector correlator $V^{(ud)}(t)$ provide us with the values of the four dimensionless parameters $R_{dual}$, $(M_\pi / E_{dual})^2$, $(M_\pi / M_\rho)^2$ and $g_{\rho \pi \pi}$, which are collected in Figs.~\ref{fig:dual_ETMC} and \ref{fig:2pion_ETMC}.
We stress that dimensionless parameters are not sensitive to the uncertainty of the scale setting.
The dependence of the four parameters on the light-quark mass $m_{ud}$, the lattice spacing $a$ and the lattice size $L$ can be described in terms of combined phenomenological fits, viz.
 \bea
      \label{eq:Rdual_fit}
      R_{dual}(m_{ud}, a^2, L) & = & R_0 \left[ 1 + R_1 m_{ud} + R_a a^2 + R_{am} a^2 m_{ud} \right] 
                                                        \left[ 1 + R_{FVE} ~ \xi \frac{e^{-M L}}{(M L)^{3/2}} \right] ~ , \\
      \label{eq:Edual_fit}
      \frac{M_\pi^2}{E_{dual}^2}(m_{ud}, a^2, L) & = & E_0 m_{ud} \left[ 1 + E_1 m_{ud} + \xi \mbox{log}(\xi) + E_2 m_{ud}^2 + E_a a^2 \right]
 \eea
and
 \bea
      \label{eq:MV_fit}
      \frac{M_\pi^2}{M_\rho^2}(m_{ud}, a^2, L) & = & V_0 m_{ud} \left[ 1 + V_1 m_{ud} + \xi \mbox{log}(\xi) + V_2 m_{ud}^2 + V_a a^2 \right]  ~ , \\
      \label{eq:gVPP_fit}
      g_{\rho \pi \pi}(m_{ud}, a^2, L) & = & g_0 \left[ 1 + g_1 m_{ud} + 2 \xi \mbox{log}(\xi) + g_a a^2 \right] 
                                                                \left[ 1 + g_{FVE} ~ \xi \frac{e^{-M L}}{(M L)^{3/2}} \right] ~ ,                                                                           
 \eea
where $M^2 \equiv 2 B_0 m_{ud}$ and $\xi \equiv M^2 / (4 \pi f_0)^2$ with $B_0$ and $f_0$ being the SU(2) low-energy constants (LECs) at LO determined in Ref.~\cite{Carrasco:2014cwa}.
Since the quantities $M_\pi^2 / E_{dual}^2$ and $M_\pi^2 / M_\rho^2$ have negligible FVEs (see the right panel of Fig.~\ref{fig:dual_ETMC} and the left panel of Fig.~\ref{fig:2pion_ETMC}), we have not included in Eqs.~(\ref{eq:Edual_fit}) and (\ref{eq:MV_fit}) any dependence on the lattice size $L$.
In Eq.~(\ref{eq:gVPP_fit}) the coefficient of the chiral log is the one predicted by ChPT at NLO~\cite{Bijnens:1998di}.
Moreover, a nonanalytic term proportional to $m_{ud}^{3/2}$ is expected from ChPT~\cite{Bijnens:1998di,Jenkins:1995vb,Aoki:2006ab} in Eqs.~(\ref{eq:MV_fit}-\ref{eq:gVPP_fit}). 
However, when we tried to include it in the fitting procedure, its coefficient was found to be well compatible with $0$.

The quality of the fits based on Eqs.~(\ref{eq:Rdual_fit}-\ref{eq:gVPP_fit}) is quite good with a $\chi^2 / d.o.f.$ always less than $1$.
All the quantities $R_{dual}$, $E_{dual}$, $M_\rho$, $g_{\rho \pi \pi}$ and $M_\pi$ are correlated with each other, since they come from fitting the ETMC vector correlators. Such correlations are properly taken into account in our bootstrap sampling procedure.
The results corresponding to the continuum and infinite volume limits are shown in Figs.~\ref{fig:dual_ETMC}-\ref{fig:2pion_ETMC} as solid lines.
In particular, at the physical pion point ($M_\pi^{phys} = M_{\pi^0} = 135$ MeV~\cite{Carrasco:2014cwa}) the value $M_\rho^{phys} \equiv M_\rho(m_{ud}^{phys}, 0, \infty) = 760 (19)$ MeV is obtained, in agreement with the experimental $\rho$-meson mass~\cite{PDG}, though within a large uncertainty.

Finally, for the simulated (squared) pion mass $M_\pi^2$ we adopt an Ansatz consistent with Eqs.~(\ref{eq:Edual_fit}-\ref{eq:MV_fit}), but including a phenomenological term for taking into account FVEs, namely
 \bea
     M_\pi^2(m_{ud}, a^2, L) & = & 2 B_0 m_{ud} \left[ 1 + P_1 m_{ud} + \xi \mbox{log}(\xi) + P_2 m_{ud}^2 + P_a a^2 \right] ~ \nonumber \\
                                            & \cdot & \left[ 1 + P_{FVE} ~ \xi \frac{e^{-M L}}{(M L)^{3/2}} \right] ~ ,
     \label{eq:MPS_fit}
 \eea
which nicely fits the lattice data and provides results consistent with those of the quark mass analysis of Ref.~\cite{Carrasco:2014cwa}.

\begin{figure}[htb!]
\centering{\scalebox{0.75}{\includegraphics{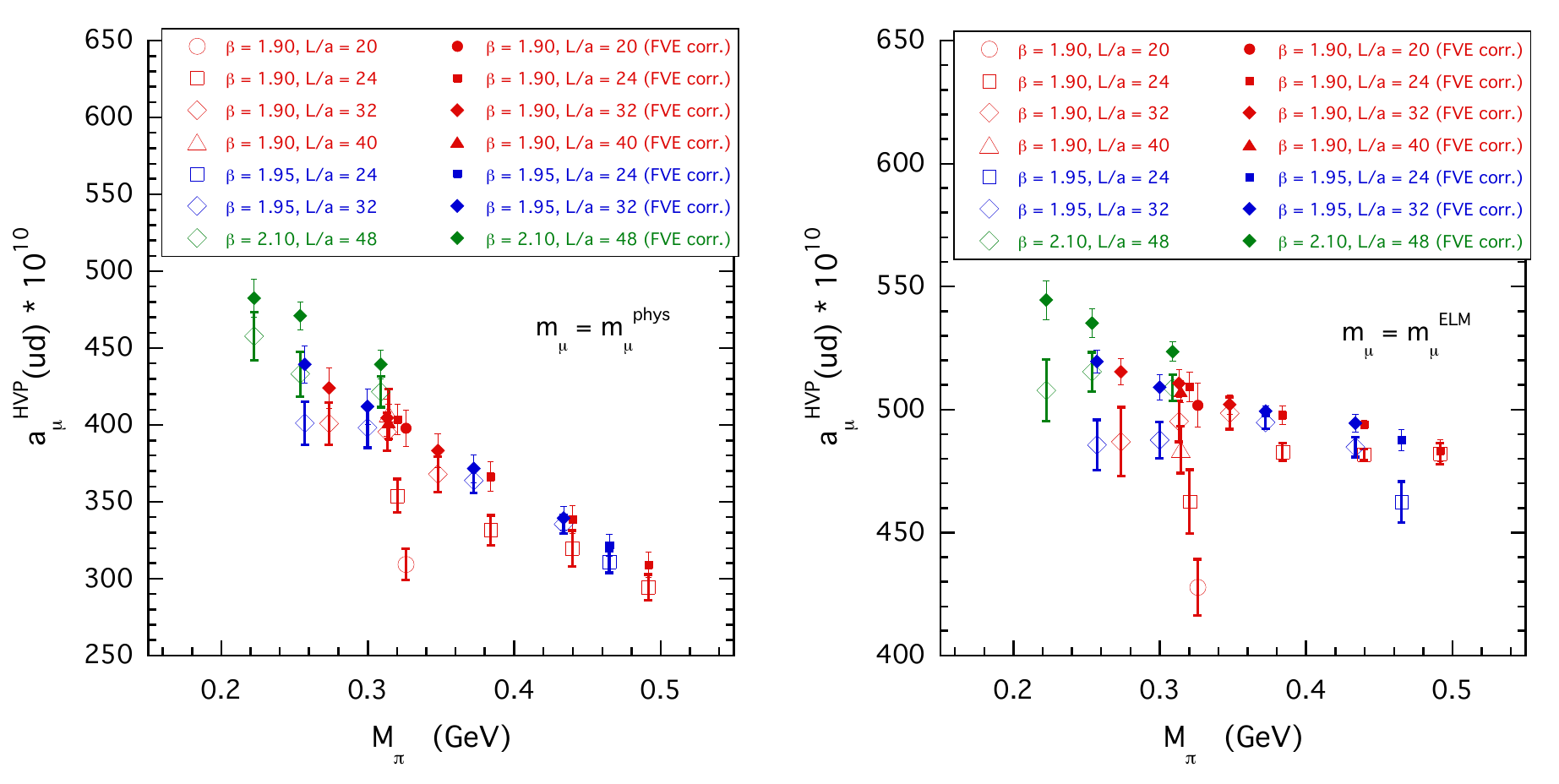}}}
\vspace{-0.25cm}
\caption{\it \small The (connected) light-quark contribution to the muon HVP, $a_\mu^{\rm HVP}(ud)$, evaluated for all the ETMC gauge ensembles of Table \ref{tab:simudetails}. Empty markers correspond to Eq.~(\ref{eq:decomposition}), where the lattice data for the vector correlator $V^{(ud)}(t)$ are directly used. Full markers are the results of the subtraction of FVEs by means of Eqs.~(\ref{eq:amud_noFVE}-\ref{eq:deltaFVE}). The physical muon mass is used in the left panel, while the ELM mass (\ref{eq:ELM}) is adopted in the right panel.\hspace*{\fill}}
\label{fig:amud_ETMC}
\end{figure}

Thus, at each value of the light-quark mass $m_{ud}$ and of the lattice spacing $a$ the fitting functions~(\ref{eq:Rdual_fit}-\ref{eq:MPS_fit}) allow us to determine the infinite volume limits  $R_{dual}(m_{ud}, a^2, \infty)$, $E_{dual}(m_{ud}, a^2, \infty)$, $M_\rho(m_{ud}, a^2, \infty)$, $g_{\rho \pi \pi}(m_{ud}, a^2, \infty)$ and $M_\pi(m_{ud}, a^2, \infty)$, which can be used in Eqs.~(\ref{eq:Vdual_vol}-\ref{eq:V2pi_vol}) to evaluate the finite-volume correction $\Delta_{FVE} a_\mu^{\rm HVP}(ud)$ for each of the ETMC ensembles.
The results of the subtraction of FVEs are illustrated in Fig.~\ref{fig:amud_ETMC}, where the (connected) light-quark contribution to the muon HVP, $a_\mu^{\rm HVP}(ud)$, is calculated using either the physical muon mass $m_\mu = m_\mu^{phys} = 105$ MeV or the effective lepton mass (ELM) $m_\mu^{ELM}$, defined as
 \be
     m_\mu^{ELM}(m_{ud}, a^2, L)  = \frac{m_\mu^{phys}}{M_\rho^{phys}} M_\rho(m_{ud}, a^2, L)  ~ .
     \label{eq:ELM}
 \ee

The ELM procedure was introduced in Ref.~\cite{Burger:2013jya} in order to weaken the light-quark mass dependence of $a_\mu^{\rm HVP}(ud)$, improving in this way the reliability of the chiral extrapolation. 
From Fig.~\ref{fig:amud_ETMC} it can be seen that the ELM procedure is able to reduce the light-quark mass dependence, but it does not modify the impact of FVEs. 
Once the latter are removed, the resulting values of $a_\mu^{\rm HVP}(ud)|_{L \to \infty}$ (see the full markers in the right panel of  Fig.~\ref{fig:amud_ETMC}) exhibit again a remarkable dependence on the light-quark mass. 

The attractive feature of the ELM procedure is based on the fact that $a_\mu^{\rm HVP}(ud)$ depends on the lepton mass in lattice units $a m_\mu$ (see Eqs.~(\ref{eq:amu_t}-\ref{eq:ftilde})). 
Thus, using Eq.~(\ref{eq:ELM}) the knowledge of the value of the lattice spacing is not required and therefore the resulting $a_\mu^{\rm HVP}(ud)$ is not affected by the uncertainties of the scale setting. 
The drawback of the ELM procedure is instead represented by its potential sensitivity to the statistical fluctuations of the $\rho$-meson mass, $a M_\rho$, determined on the lattice.

We close this section by observing that:
\begin{itemize}

 \item the use of the analytic representation (\ref{eq:dual+2pi}) of the vector correlator $V^{(ud)}(t)$ allows to subtract the FVEs on $a_\mu^{\rm HVP}(ud)$ relying only on lattice data;

\item the light-quark mass dependence of $a_\mu^{\rm HVP}(ud)$ becomes remarkably steeper after the subtraction of FVEs, which means that any reliable chiral extrapolation or interpolation of the lattice values of $a_\mu^{\rm HVP}(ud)$ cannot be carried out without taking care of FVEs properly.

\end{itemize}

\section{Extrapolations to the physical pion point and to the continuum limit}
\label{sec:extrapolations}

In this section we perform the extrapolation to the physical pion point and to the continuum limit of the lattice data corrected by the FVEs as discussed in the previous section (see the full markers in Fig.~\ref{fig:amud_ETMC}). 
An important feature of the chiral behavior of $a_\mu^{\rm HVP}(ud)$ is that it diverges in the chiral limit $m_{ud} \to 0$~\cite{Golowich:1995kd,Amoros:1999dp,Bijnens:2016ndo}.
This is connected with the loss of analyticity of the (subtracted) HVP function at vanishing photon virtuality $Q^2 = 0$ in that limit.
As a consequence, ChPT predicts already at NLO the presence of a chiral log proportional to $\mbox{log}(m_{ud})$~\cite{Golterman:2017njs}.

The ChPT expansion can be applied to the HVP form factor $\Pi_R^{(ud)}(Q^2)$ appearing in the covariant decomposition of the HVP tensor related to the $u$- and $d$-quark em~currents~\cite{Golowich:1995kd,Amoros:1999dp,Bijnens:2016ndo,Golterman:2017njs}.
For the connected part of $\Pi_R^{(ud)}(Q^2)$ one has
 \be
      \Pi_R^{(ud)}(Q^2) = \frac{5}{9} \left[ \Pi_R^{\rm NLO}(Q^2) + \Pi_R^{\rm NNLO}(Q^2) + ... \right]
      \label{eq:PiR_ChPT}
 \ee
with
 \bea
       \label{eq:PiR_NLO}
       \Pi_R^{\rm NLO}(Q^2) & = & \frac{1}{24 \pi^2} \left[ 2 \widehat{B}(Q^2, M_\pi^2) + \widehat{B}(Q^2, M_K^2) \right] ~ , \\
        \Pi_R^{\rm NNLO}(Q^2) & = & \frac{1}{72 \pi^2} \frac{Q^2}{16 \pi^2 f_\pi^2} \left[ 2 B(Q^2, M_\pi^2) + B(Q^2, M_K^2) \right]^2  ~ \nonumber \\
                                               & - & \frac{16}{3} L_9^r(\mu_\chi) \frac{Q^2}{16 \pi^2 f_\pi^2} \left[ 2 B(Q^2, M_\pi^2) + B(Q^2, M_K^2) \right] - 8 C_{93}^r(\mu_\chi) Q^2 ~ ,
       \label{eq:PiR_NNLO}
 \eea 
where $\mu_\chi$ is the ChPT renormalization scale and
 \bea
        \label{eq:ChPT_B}
       B(Q^2, M^2) & \equiv & \frac{1}{2} \left[ 1 + \mbox{log}\left( \frac{M^2}{\mu_\chi^2} \right) \right] + \widehat{B}(Q^2, M^2) ~ , \\
        \label{eq:ChPT_hatB}
       \widehat{B}(Q^2, M^2) & = & \widehat{B}\left( x = \frac{4M^2}{Q^2} \right) = (1 + x)^{3/2} ~ \mbox{log}\left( \frac{1 + \sqrt{1 + x}}{\sqrt{x}} \right) -x - \frac{4}{3} ~ .
 \eea
The NLO term (\ref{eq:PiR_NLO}) is independent of any LECs, while at NNLO two LECs appear in Eq.~(\ref{eq:PiR_NNLO}), namely $L_9^r(\mu_\chi)$ and $C_{93}^r(\mu_\chi)$.

The NLO and NNLO contributions to $a_\mu^{\rm HVP}(ud)$ can be evaluated using the following expression  
\bea
      \label{eq:amud_ChPT}
      \left[ a_\mu^{\rm HVP}(ud) \right]^{\rm NLO(NNLO)} = \frac{40}{9} \alpha_{em}^2 \int_0^\infty dz && ~ \frac{1}{\sqrt{4 + z^2}} ~ 
                                                                                            \left( \frac{\sqrt{4 + z^2} - z}{\sqrt{4 + z^2} + z} \right)^2 \nonumber \\
                                                                                            & \cdot & \Pi_R^{\rm NLO(NNLO)}(m_\mu^2 z^2) ~ .
\eea

\begin{figure}[htb!]
\centering{\scalebox{0.75}{\includegraphics{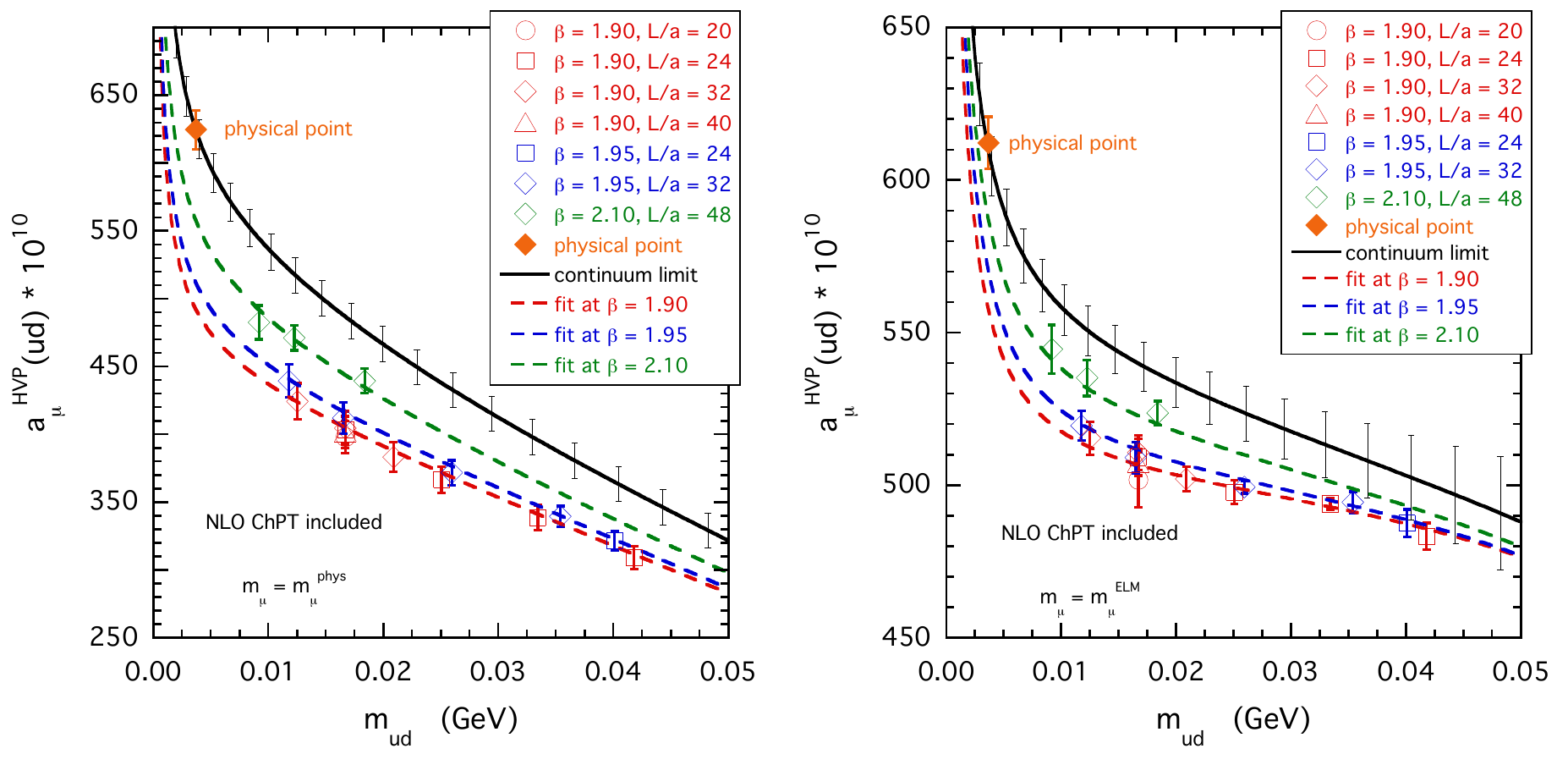}}}
\vspace{-0.25cm}
\caption{\it \small Values of the (connected) light-quark contribution to the muon HVP, $a_\mu^{\rm HVP}(ud)$, corrected by FVEs and evaluated using either $m_\mu = m_\mu^{phys} = 105$ MeV (left panel) or $m_\mu = m_\mu^{ELM}$ (right panel). The dashed lines represent the fitting function (\ref{eq:NLOfit}), which includes the NLO ChPT prediction, evaluated at each value of the lattice spacing of the ETMC ensembles. The solid lines represent the same fitting function in the continuum limit. The full (orange) diamonds are the values extrapolated at the physical pion point and in the continuum limit.\hspace*{\fill}}
\label{fig:NLOfit}
\end{figure}

\begin{figure}[htb!]
\centering{\scalebox{0.75}{\includegraphics{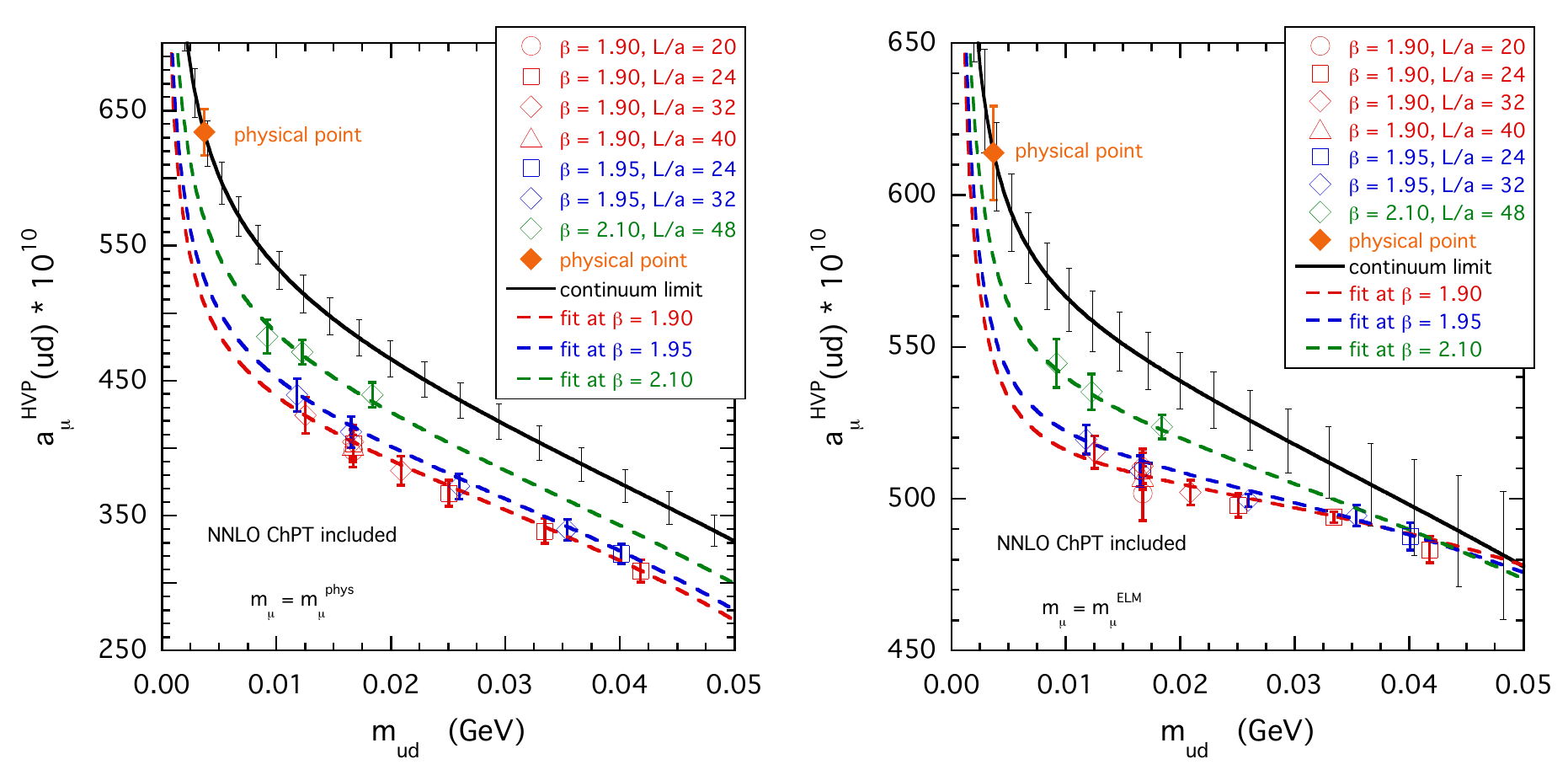}}}
\vspace{-0.25cm}
\caption{\it \small The same as in Fig.~\ref{fig:NLOfit}, but adopting the fitting function (\ref{eq:NNLOfit}), which includes also the NNLO ChPT prediction.\hspace*{\fill}}
\label{fig:NNLOfit}
\end{figure}

Thus, we have adopted three different fitting functions, which, besides discretization effects, include in different ways the effects of chiral logs, namely
\begin{itemize}
\item including NLO ChPT:
\be
      \label{eq:NLOfit}
      a_\mu^{\rm HVP}(ud) = \left\{ \left[ a_\mu^{\rm HVP}(ud) \right]^{\rm NLO} + A_0 + A_1 m_{ud} + A_2 m_{ud}^2 \right\} \cdot
                                            \left[ 1 + D_0 a^2 + D_1 a^2 m_{ud} \right] ~ ,
\ee

\item including NLO and NNLO ChPT:
\bea
       \label{eq:NNLOfit}
      a_\mu^{\rm HVP}(ud) & = & \left\{ \left[ a_\mu^{\rm HVP}(ud) \right]^{\rm NLO} + \left[ a_\mu^{\rm HVP}(ud) \right]^{\rm NNLO} + A_0^\prime + 
                                                   A_1^\prime m_{ud} \right\} \nonumber \\
                                         & \cdot & \left[ 1 + D_0^\prime a^2 + D_1^\prime a^2 m_{ud} \right] ~ ,
\eea

\item including free logs:
\bea
      \label{eq:freelogfit}
      a_\mu^{\rm HVP}(ud) & = & \left( \widetilde{A}_0 + \widetilde{A}_0^{log} \mbox{log}(m_{ud}) \right) \left( 1 + \widetilde{A}_1 m_{ud} + 
                                                    \widetilde{A}_1^{log} m_{ud} \mbox{log}(m_{ud}) \right) \nonumber \\
                                         & \cdot & \left[ 1 + \widetilde{D}_0 a^2 + \widetilde{D}_1 a^2 m_{ud} \right] ~ ,
\eea
\end{itemize}
where, for the sake of simplicity, $a_\mu^{\rm HVP}(ud)$ stands from now on for $a_\mu^{\rm HVP}(ud)|_{L \to \infty}$ (see Eqs.~(\ref{eq:amud_noFVE}-\ref{eq:deltaFVE})).
The results of the combined chiral extrapolation and continuum limit obtained using either Eq.~(\ref{eq:NLOfit}) or Eq.~(\ref{eq:NNLOfit}) are shown in Figs.~\ref{fig:NLOfit} and \ref{fig:NNLOfit}, respectively, with and without the use of the ELM procedure.
Similar results are obtained in the case of the "free logs" fitting function (\ref{eq:freelogfit}).

In the case of the NNLO fitting function (\ref{eq:NNLOfit}) we get the following values for the LECs $L_9^r$ and $C_{93}^r$ at the $\rho$-meson mass scale $\mu_\chi = 0.77$ GeV:
\bea
      L_9^r(0.77~\mbox{GeV}) & = & 0.00273 ~ (143) ~ , \\
      C_{93}^r(0.77~\mbox{GeV}) & = & -0.0136 ~ (20) ~ \mbox{GeV}^{-2} ~ ,
 \eea
which are consistent (within the uncertainties) with the findings $L_9^r(0.77~\mbox{GeV}) = 0.00593 ~ (43)$ and $C_{93}^r(0.77~\mbox{GeV}) = -0.0154 ~ (4)$ GeV$^{-2}$ obtained in Ref.~\cite{Golterman:2017njs}.

\begin{figure}[htb!]
\centering{\scalebox{0.75}{\includegraphics{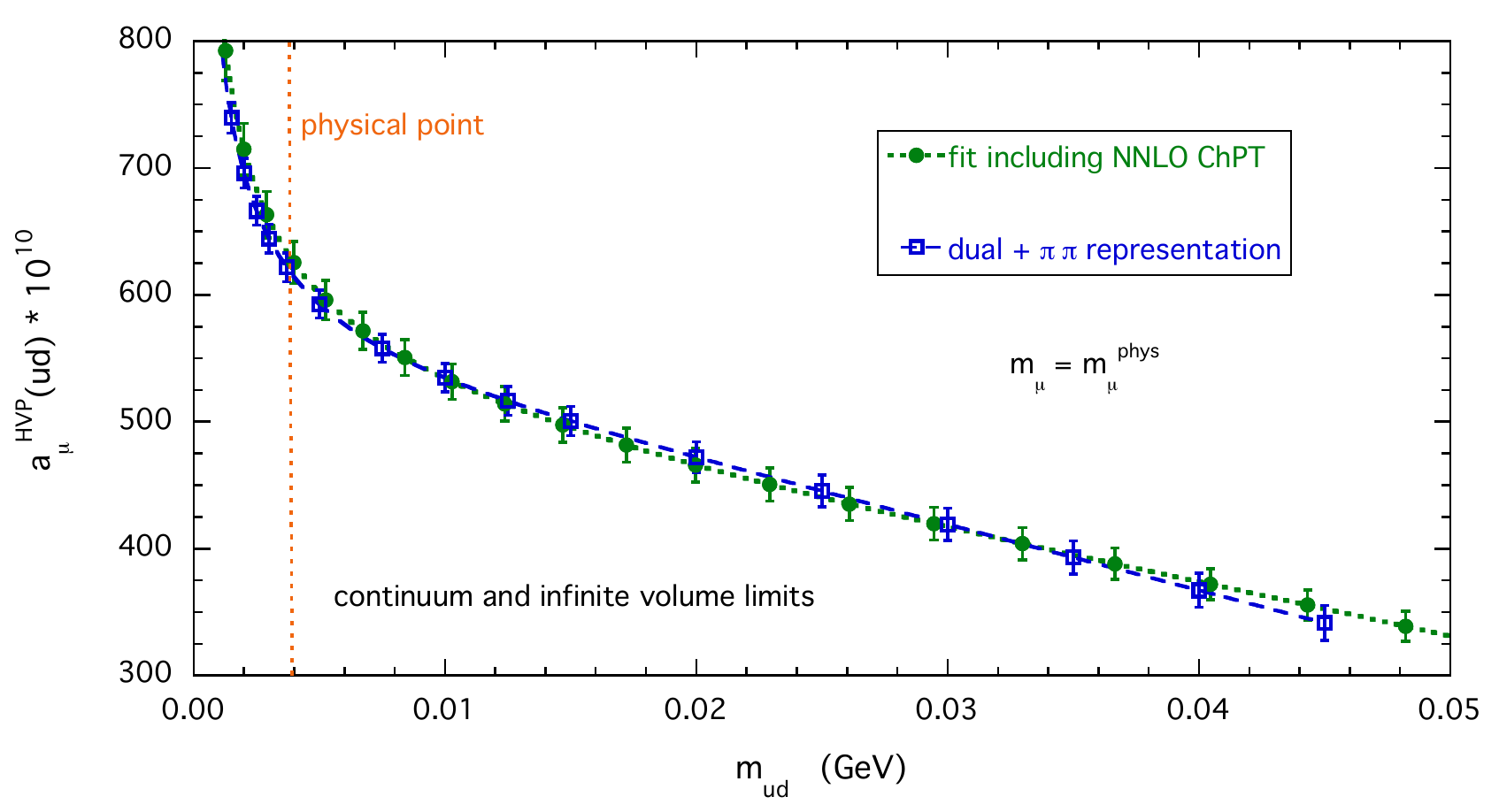}}}
\vspace{-0.25cm}
\caption{\it \small The light-quark mass dependence of $a_\mu^{\rm HVP}(ud)$, in units of $10^{10}$, obtained in the continuum and infinite volume limits using the physical muon mass. The blue squares correspond to the predictions of our ``dual + $\pi \pi$" analytic representation (\ref{eq:Vdual_vol}-\ref{eq:V2pi_vol}) evaluated using the values $R_{dual}(m_{ud}, 0, \infty)$, $E_{dual}(m_{ud}, 0, \infty)$, $M_\rho(m_{ud}, 0, \infty)$, $g_{\rho \pi \pi}(m_{ud}, 0, \infty)$ and $M_\pi(m_{ud}, 0, \infty)$ obtained from Eqs.~(\ref{eq:Rdual_fit}-\ref{eq:MPS_fit}). The green dots represent the results of the ChPT fit (\ref{eq:NNLOfit}) taken in the continuum limit.\hspace*{\fill}}
\label{fig:amud_rep}
\end{figure}

From Figs.~\ref{fig:NLOfit} and \ref{fig:NNLOfit} it can be clearly seen that the enhancement due to chiral logs is important close to the physical point.
This makes $a_\mu^{\rm HVP}(ud)$ quite sensitive to small changes of the light-quark mass, which may be crucial even for a local interpolation around the physical point.
This immediately rises the question of how much we can trust the sharp rise visible in Figs.~\ref{fig:NLOfit} and \ref{fig:NNLOfit}. 
In order to address this issue we resort to our ``dual + $\pi \pi$" analytic representation.
At each value of the light-quark mass $m_{ud}$ we can determine the values $R_{dual}(m_{ud}, 0, \infty)$, $E_{dual}(m_{ud}, 0, \infty)$, $M_\rho(m_{ud}, 0, \infty)$, $g_{\rho \pi \pi}(m_{ud}, 0, \infty)$ and $M_\pi(m_{ud}, 0, \infty)$ from the fitting functions~(\ref{eq:Rdual_fit}-\ref{eq:MPS_fit}) (i.e.~the solid lines in Figs.~\ref{fig:dual_ETMC} and \ref{fig:2pion_ETMC}).
Then, by means of Eqs.~(\ref{eq:Vdual_vol}-\ref{eq:V2pi_vol}) we estimate the light-quark mass dependence of $a_\mu^{\rm HVP}(ud)$.
The corresponding results are shown in Fig.~\ref{fig:amud_rep} by the blue squares and compared with those obtained using the fitting function (\ref{eq:NNLOfit}) in the continuum limit (green dots).
A remarkable agreement is observed not only at large values of $m_{ud}$ (where our analytic representation fits very nicely the ETMC vector correlators), but also at values of $m_{ud}$ close and even smaller than the physical point. 
We point out that our ``dual + $\pi \pi$" analytic representation does not contain chiral logs explicitly and, therefore, the agreement with the ChPT fit shown in Fig.~\ref{fig:amud_rep} is reassuring about the reliability of the sharp rise of $a_\mu^{\rm HVP}(ud)$ at low values of the light-quark mass $m_{ud}$.

The results obtained at the physical pion point and in the continuum (and infinite volume) limit using the fitting functions (\ref{eq:NLOfit}-\ref{eq:freelogfit}) and adopting for the muon mass either its physical value ($m_\mu^{phys} = 105$ MeV) or the ELM value (\ref{eq:ELM}) are collected in Table~\ref{tab:amud_phys}.
 \begin{table}[hbt!]
\begin{center}
\renewcommand{\arraystretch}{1.10}
\begin{tabular}{||c||c|c|c||}
\hline
  & with NLO ChPT & with NNLO ChPT & ~~ free logs ~~ \\
  & Eq.~(\ref{eq:NLOfit}) & Eq.~(\ref{eq:NNLOfit}) & Eq.~(\ref{eq:freelogfit}) \\ 
\hline \hline
$m_\mu = m_\mu^{phys}$ & 624.6 (14.5) & 634.1 (17.0) & 613.1 (13.2) \\
\hline
$m_\mu = m_\mu^{ELM}$ & 612.3~~(8.5) & 613.8 (15.4) & 616.4 (17.5) \\
\hline \hline
\end{tabular}
\renewcommand{\arraystretch}{1.0}
\end{center}
\vspace{-0.25cm}
\caption{\it \small Values of $a_\mu^{\rm HVP}(ud)$, in units of $10^{10}$, extrapolated to the physical pion point and to the continuum limit using the fitting functions (\ref{eq:NLOfit}-\ref{eq:freelogfit}) and adopting for the muon mass either its physical value ($m_\mu^{phys} = 105$ MeV) or the ELM value (\ref{eq:ELM}).\hspace*{\fill}}
\label{tab:amud_phys}
\end{table}
Using the averaging procedure given by Eq.~(28) of Ref.~\cite{Carrasco:2014cwa} we get
 \bea
      a_\mu^{\rm HVP}(ud) & = & 619.0 ~ (14.7)_{stat+fit+input} ~ (6.2)_{chir} ~ (4.9)_{disc} ~ (6.2)_{FVE} \cdot 10^{-10} \nonumber \\
                                         & = & 619.0 ~ (17.8) \cdot 10^{-10} ~ ,
      \label{eq:amud_final}
 \eea
where
\begin{itemize}
\item $()_{stat+fit+input}$ incorporates the uncertainties induced by both the statistical errors and the fitting procedure itself as well as the error coming from the uncertainties of the input parameters of the eight branches of the quark mass analysis of Ref.~\cite{Carrasco:2014cwa};
\item $()_{chir}$ is the error due to the chiral extrapolation estimated from the spread of the results corresponding to the three fitting functions (\ref{eq:NLOfit}-\ref{eq:freelogfit});
\item $()_{disc}$ is the uncertainty due to both discretization effects and scale setting, estimated by comparing the results obtained with and without the ELM procedure (\ref{eq:ELM});
\item $()_{FVE}$ is the error due to the subtraction of FVEs, taken conservatively to be twice the uncertainty found in subsection~\ref{sec:FVE_A40XX} (see Table~\ref{tab:Fpi_parm}).
\end{itemize}

Our finding (\ref{eq:amud_final}) improves the previous ETMC estimate of Ref.~\cite{Burger:2013jya}, $a_\mu^{\rm HVP}(ud) = 567 ~ (11) \cdot 10^{-10}$, thanks to a  more accurate treatment of both the FVEs and the extrapolation to the physical pion point.
The latter can be clearly avoided using ensembles close to the physical point.
Recently ETMC has generated a gauge ensemble close to the physical pion mass with $N_f = 2$ dynamical quarks, obtaining the value $a_\mu^{\rm HVP}(ud) = 552 ~ (39) \cdot 10^{-10}$~\cite{Abdel-Rehim:2015pwa}.
The lattice size is $L \simeq 4.4$ fm, which corresponds to $M_\pi L \simeq 3.0$.
For such a setup we expect large FVEs, which will be discussed in the next Section (see later Fig.~\ref{fig:FVEs}).
For the setup chosen in Ref.~\cite{Abdel-Rehim:2015pwa} we estimate a correction due to FVEs of order of $10 \%$, which would yield a final value $a_\mu^{\rm HVP}(ud) \simeq 610 ~ (40) \cdot 10^{-10}$ in agreement with Eq.~(\ref{eq:amud_final}), though within a large uncertainty.

\begin{figure}[htb!]
\centering{\scalebox{0.75}{\includegraphics{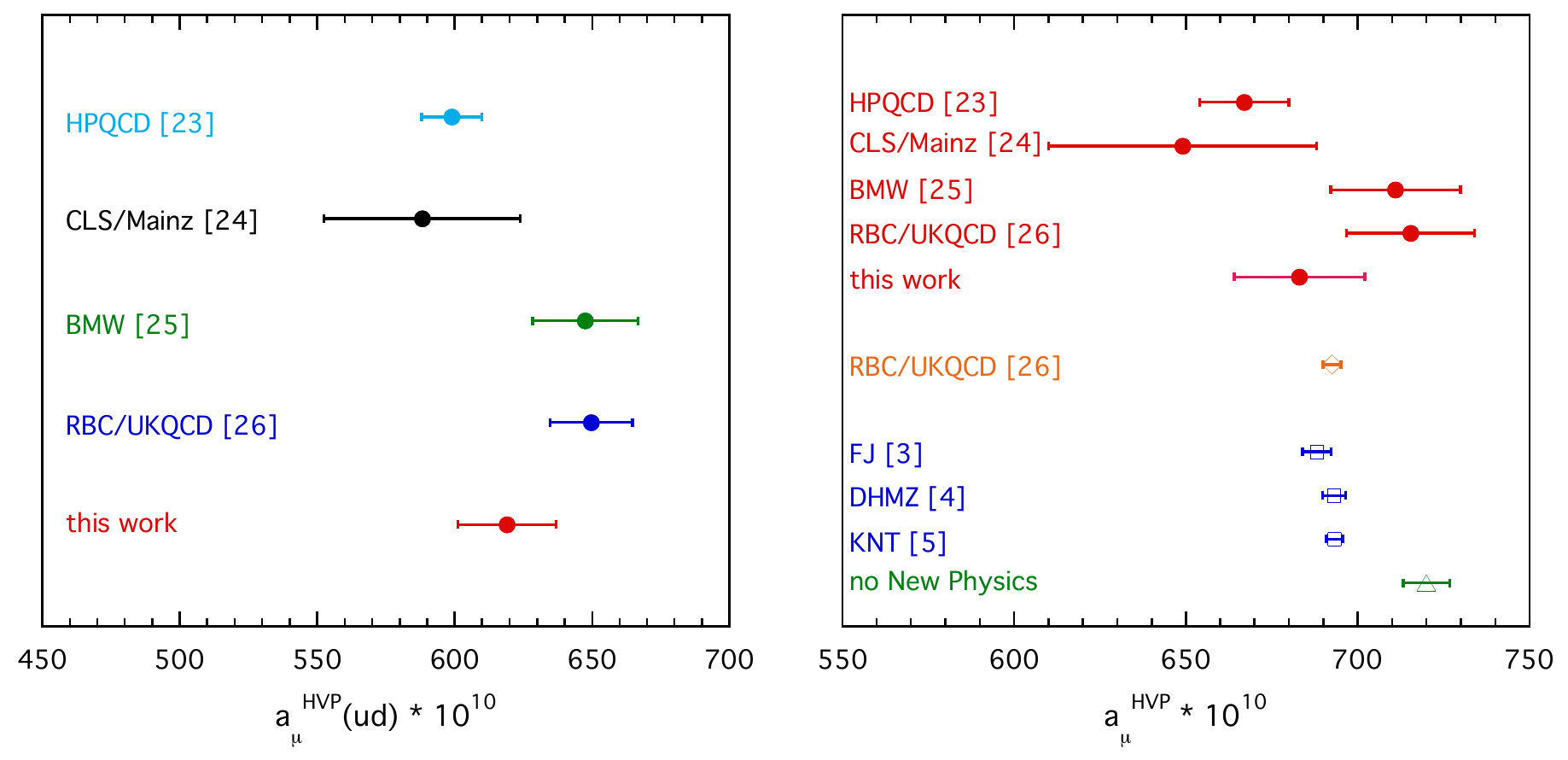}}}
\vspace{-0.25cm}
\caption{\it \small Left panel: values of the connected light-quark contribution to the muon HVP, $a_\mu^{\rm HVP}(ud)$, obtained at the physical pion point and in the continuum and infinite volume limits in the present work (\ref{eq:amud_final}), and by HPQCD~\cite{Chakraborty:2016mwy}, CLS/Mainz~\cite{DellaMorte:2017dyu}, BMW~\cite{Borsanyi:2017zdw} and RBC/UKQCD~\cite{Blum:2018mom}. Right panel: values of the muon HVP $a_\mu^{\rm HVP}(udsc)$ obtained in the present work (\ref{eq:amu_final}) and in Refs.~\cite{Chakraborty:2016mwy,DellaMorte:2017dyu,Borsanyi:2017zdw,Blum:2018mom}. The result obtained in Ref.~\cite{Blum:2018mom} using the R-ratio method (which turns out to be based on lattice points by $\simeq 30\%$ and on dispersive $e^+ e^-$ data by $\simeq 70 \%$) is also included as an orange dot. The results of the recent dispersive analysis of Refs.~\cite{Jegerlehner:2017lbd,Davier:2017zfy,Keshavarzi:2018mgv} are shown together with the value of $a_\mu^{HV}$ corresponding to a vanishing muon anomaly (labeled as ``no New Physics'').\hspace*{\fill}}
\label{fig:amu_final}
\end{figure}

Our result (\ref{eq:amud_final}) is compared with the most recent ones from other lattice collaborations in the left panel of Fig.~\ref{fig:amu_final}.
Within the errors our value obtained with $N_f = 2+1+1$ dynamical flavors of sea quarks agrees with the corresponding results from HPQCD~\cite{Chakraborty:2016mwy} ($N_f = 2+1+1$), CLS/Mainz~\cite{DellaMorte:2017dyu} ($N_f = 2$), BMW~\cite{Borsanyi:2017zdw} ($N_f = 2+1+1$) and RBC/UKQCD~\cite{Blum:2018mom} ($N_f = 2+1$).

Adding the connected contributions from strange and charm quarks, $a_\mu^{\rm HVP}(s) = 53.1 ~ (2.5) \cdot 10^{-10}$ and $a_\mu^{\rm HVP}(c) = 14.75 ~ (0.56) \cdot 10^{-10}$ determined by ETMC in Ref.~\cite{Giusti:2017jof}, and an estimate of the IB corrections $a_\mu^{\rm HVP}(IB) = 8 ~ (5) \cdot 10^{-10}$ and of the quark disconnected diagrams $a_\mu^{\rm HVP}(disconn.) = -12 ~ (4) \cdot 10^{-10}$, obtained using the findings of Refs.~\cite{Borsanyi:2017zdw} and \cite{Blum:2018mom}, we finally get for the muon HVP $a_\mu^{\rm HVP}(udsc)$ the value
 \be
       a_\mu^{\rm HVP}(udsc) = 683 ~ (19) \cdot 10^{-10} ~ ,
       \label{eq:amu_final}
 \ee
which is in nice agreement with the recent results $a_\mu^{\rm HVP} = 688.07 ~ (4.14) \cdot 10^{-10}$~\cite{Jegerlehner:2017lbd}, $a_\mu^{\rm HVP} = 693.10 ~ (3.40) \cdot 10^{-10}$~\cite{Davier:2017zfy} and $a_\mu^{\rm HVP} = 693.27 ~ (2.46) \cdot 10^{-10}$~\cite{Keshavarzi:2018mgv}, based on dispersive analyses of the experimental cross section data for $e^+ e^-$ annihilation into hadrons. 
Our value (\ref{eq:amu_final}) is compared with the results of other lattice collaborations as well as with the dispersive results of Refs.~\cite{Jegerlehner:2017lbd,Davier:2017zfy,Keshavarzi:2018mgv} in the right panel of Fig.~\ref{fig:amu_final}.

\section{Light-quark vector correlator at the physical pion point and moments of the polarization function}
\label{sec:physical}

In this section we apply our analytic representation (\ref{eq:dual+2pi}) to estimate the connected light-quark vector correlator $V^{(ud)}(t)$ at the physical pion point both for finite values of the lattice size $L$ and in the infinite volume limit.

To this end at each value of the lattice size $L$ we determine the values $R_{dual}(m_{ud}^{phys}, 0, L)$, $E_{dual}(m_{ud}^{phys}, 0, L)$, $M_\rho(m_{ud}^{phys}, 0, L)$, $g_{\rho \pi \pi}(m_{ud}^{phys}, 0, L)$ and $M_\pi(m_{ud}^{phys}, 0, L)$ from the fitting functions~(\ref{eq:Rdual_fit}-\ref{eq:MPS_fit}), where $m_{ud}^{phys} = 3.70 ~ (17)$ MeV as determined in Ref.~\cite{Carrasco:2014cwa}.
We use the above values in Eqs.~(\ref{eq:V2pi}) and (\ref{eq:Vdual}) to obtain the connected light-quark vector correlator $V^{(ud)}(t)$ at the physical pion point and at finite values of $L$.

The infinite volume limit is constructed by determining the values $R_{dual}(m_{ud}^{phys}, 0, \infty) = 1.14 ~ (6)$, $E_{dual}(m_{ud}^{phys}, 0, \infty) = 479 ~ (22)$ MeV, $M_\rho(m_{ud}^{phys}, 0, \infty) = 760 ~ (19)$ MeV, $g_{\rho \pi \pi}(m_{ud}^{phys}, 0, \infty) = 5.30 ~ (5)$ and $M_\pi(m_{ud}^{phys}, 0, \infty) = 135$ MeV from the fitting functions~(\ref{eq:Rdual_fit}-\ref{eq:MPS_fit}).
The above values are used in Eqs.~(\ref{eq:Vdual_vol}) and (\ref{eq:V2pi_vol}) to get the connected light-quark vector correlator $V^{(ud)}(t)$ at the physical pion point and in the infinite volume limit.

\begin{figure}[htb!]
\centering{\scalebox{0.75}{\includegraphics{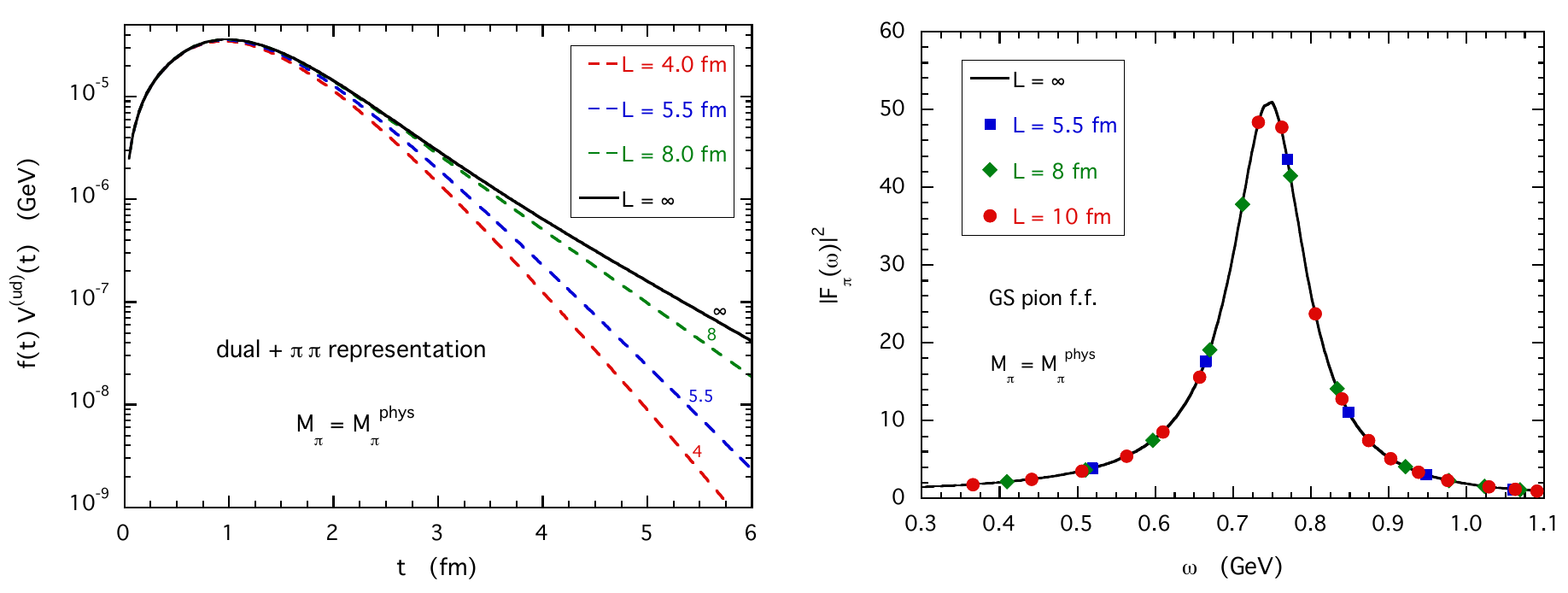}}}
\vspace{-0.25cm}
\caption{\it \small Left panel: light-quark vector correlator $V^{(ud)}(t)$, multiplied by the muon kernel $f(t)$, evaluated using our ``dual + $\pi \pi$" representation (\ref{eq:dual+2pi}) extrapolated at the physical pion point and in the continuum limit for three values of the lattice size $L$ (see text). The infinite volume limit, constructed as explained in the text, is also shown by the black solid line. Right panel: the (squared) pion form factor $|F_\pi(\omega)|^2$ corresponding to the GS parametrization (\ref{eq:Fpi_GS}-\ref{eq:Apipi0}) evaluated in the infinite volume limit (see text) versus the two-pion energy $\omega$. The full dots are located at the position of the energy levels satisfying the L\"uscher condition (\ref{eq:kn}) for each value of the lattice size $L$.\hspace*{\fill}}
\label{fig:pionpoint}
\end{figure}

The results obtained for few values of the lattice size $L$ and in the infinite volume limit are shown in the left panel of Fig.~\ref{fig:pionpoint}.
The number of elastic energy levels included in Eq.~(\ref{eq:V2pi}) depends on $L$ and, at the physical pion point, it is larger than $4$, i.e.~of the number of states used in the fitting procedure of the ETMC vector correlators.
The right panel of Fig.~\ref{fig:pionpoint} illustrates this point.
There, the full dots represent the position of the energy levels satisfying the L\"uscher condition (\ref{eq:kn}) for few values of $L$ and, at the same time, the values of the (squared) GS pion form factor occurring in Eq.~(\ref{eq:An}).

We observe that from the threshold up to $\omega \sim 1 $ GeV the number of energy levels is $5$ for $L = 5.5$ fm, $8$ for $L = 8$ fm and reaches $14$ for $L = 10$ fm.
Therefore, at the physical pion point the spectral decomposition of the vector correlator $V^{(ud)}(t)$ is quite involved.
Very large time distances should be reached for getting the dominance of the lowest energy level, because the corresponding coupling $A_{n =1}$ is quite small.
Higher energy levels fall off faster, but they have larger values of the coupling $A_n$ up to the location of the $\rho$-meson resonance.
The consequences are that: ~ i) the FVEs on the tail of $V^{(ud)}(t)$ increase significantly as the time distance increases, and ~ ii) the effective mass of the light-quark vector correlator (see Eq.~(\ref{eq:Meff})) does not show any plateau for time distances currently accessible on the lattice.

\begin{figure}[htb!]
\centering{\scalebox{0.75}{\includegraphics{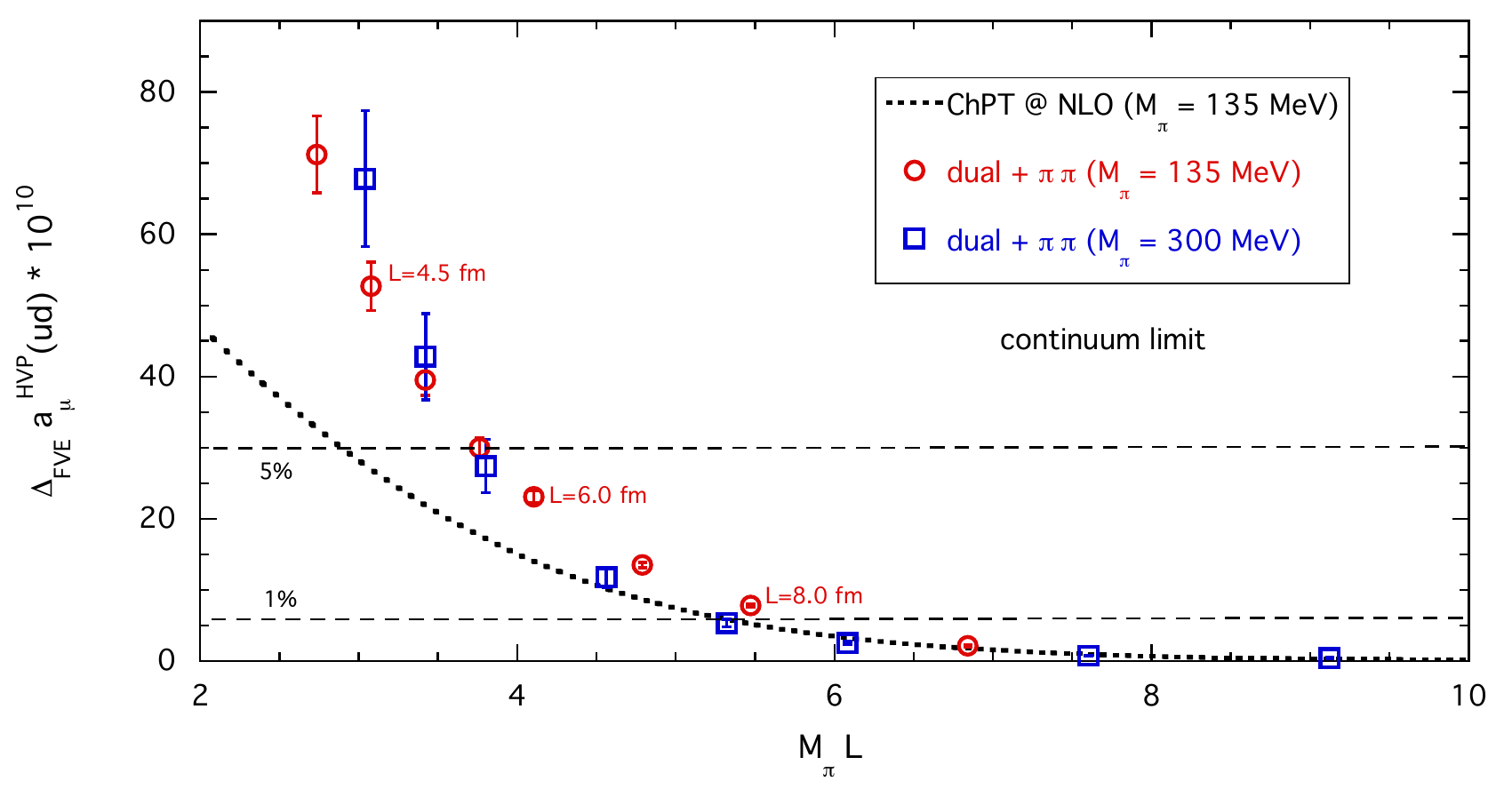}}}
\vspace{-0.25cm}
\caption{\it \small Values of $\Delta_{FVE} a_\mu^{\rm HVP}(ud)$ (see Eq.~(\ref{eq:deltaFVE})), evaluated in the continuum limit according to our ``dual + $\pi \pi$" representation at the physical pion point (red circles) and at a larger pion mass equal to $M_\pi = 300$ MeV (blue squares). The dotted line corresponds to the predictions of ChPT at NLO~\cite{Aubin:2015rzx,Bijnens:2017esv}.\hspace*{\fill}}
\label{fig:FVEs}
\end{figure}

In Fig.~\ref{fig:FVEs} the FVEs estimated at the physical pion mass on $a_\mu^{HVP}(ud)$ by means of Eq.~(\ref{eq:deltaFVE}) are shown versus $M_\pi L$ and compared with the predictions of ChPT at NLO~\cite{Aubin:2015rzx,Bijnens:2017esv}.
The latter ones coincide with the FVEs corresponding to noninteracting two-pion states~\cite{Francis:2013qna,DellaMorte:2017dyu}.
Our determination of FVEs, instead, takes into account the interaction in the two-pion system, and in particular the resonant scattering between two pions in P-wave with total isospin 1. 
Our estimate of FVEs is significantly larger than the ChPT NLO prediction.
Recently, FVEs in the polarization function close to the physical pion point have been analyzed in ChPT at NNLO~\cite{Bijnens:2017esv}, but the corresponding numerical findings seem to be too small to explain the differences with our determination.

In Fig.~\ref{fig:FVEs} we have also shown the results for $\Delta_{FVE} a_\mu^{\rm HVP}(ud)$ at a larger pion mass equal to $M_\pi = 300$ MeV. 
At fixed values of $M_\pi L$ the FVEs on $a_\mu^{HVP}(ud)$ appear to be only slightly dependent on the pion mass (at variance with what occurs in case of the pion mass and decay constant).

At the physical pion point FVEs of the order of the muon anomaly (i.e., $\simeq 5\%$) are expected to occur for $L \simeq 5.5$ fm.
In order to reach a finite volume correction of the order of $\simeq 1\%$ or less a lattice size $L$ larger than $\simeq 8$ fm is required.

Recently, in Ref.~\cite{Borsanyi:2016lpl} the slope and the curvature of the leading HVP function at vanishing photon virtuality have been determined on the lattice at the physical pion point and in the continuum and infinite volume limits.
These quantities are derivatives of the HVP function evaluated at $Q^2 = 0$ and they can be easily related to time-moments of the vector correlator.
The separate contributions arising from the (connected) light, strange and charm quarks are also provided in Ref.~\cite{Borsanyi:2016lpl}.
Thus, for a comparison with the predictions of our ``dual + $\pi \pi$" representation of the vector correlator $V^{(ud)}(t)$ we consider the following time moments 
 \be
      \label{eq:Pin}
      \Pi_{n + 1}^{(ud)} \equiv (-)^n \frac{(n + 1)!}{(2n + 4)!} ~ \frac{18}{5} \int_0^\infty dt ~ t^{2n + 4} V^{(ud)}(t)
 \ee 
with $n = 0, 1, 2, ...$.
The quantities $\Pi_1^{(ud)}$ and $\Pi_2^{(ud)}$ correspond respectively to the slope and the curvature determined in Ref.~\cite{Borsanyi:2016lpl}.
There, it has been shown that the time distances that need to be reached to reliably determine the slope and the curvature are above $\sim 2$ and $\sim 4$ fm, respectively. 
At the physical pion point and in the continuum and infinite volume limits the predictions of our ``dual + $\pi \pi$" representation are
 \be
      \Pi_1^{(ud)} = 0.1642 ~ (33) ~ \mbox{GeV}^{-2} ~ , \qquad \qquad  \Pi_2^{(ud)} = - 0.383 ~ (16) ~ \mbox{GeV}^{-4} ~ ,
 \ee
which can be compared with the results $\Pi_1^{(ud)} = 0.1659 ~ (33)$ GeV$^{-2}$ and $\Pi_2^{(ud)} = - 0.311 ~ (16)$ GeV$^{-2}$ from Ref.~\cite{Borsanyi:2016lpl}.
The agreement is quite good in the case of the slope, while our curvature is (in absolute value) larger than the corresponding result of Ref.~\cite{Borsanyi:2016lpl} by $\simeq 20 \%$.
We note that in Ref.~\cite{Borsanyi:2016lpl} FVEs are estimated using ChPT at NLO and, therefore, the difference with our result is likely to be ascribed to the treatment of FVEs.

In the case of the higher moments $\Pi_3^{(ud)}$ and $\Pi_4^{(ud)}$ our results are
 \be
      \Pi_3^{(ud)} = 1.394 ~ (65) ~ \mbox{GeV}^{-6} ~ , \qquad \qquad  \Pi_4^{(ud)} = - 7.60 ~ (28) ~ \mbox{GeV}^{-8} ~ .
 \ee

\begin{figure}[htb!]
\centering{\scalebox{0.75}{\includegraphics{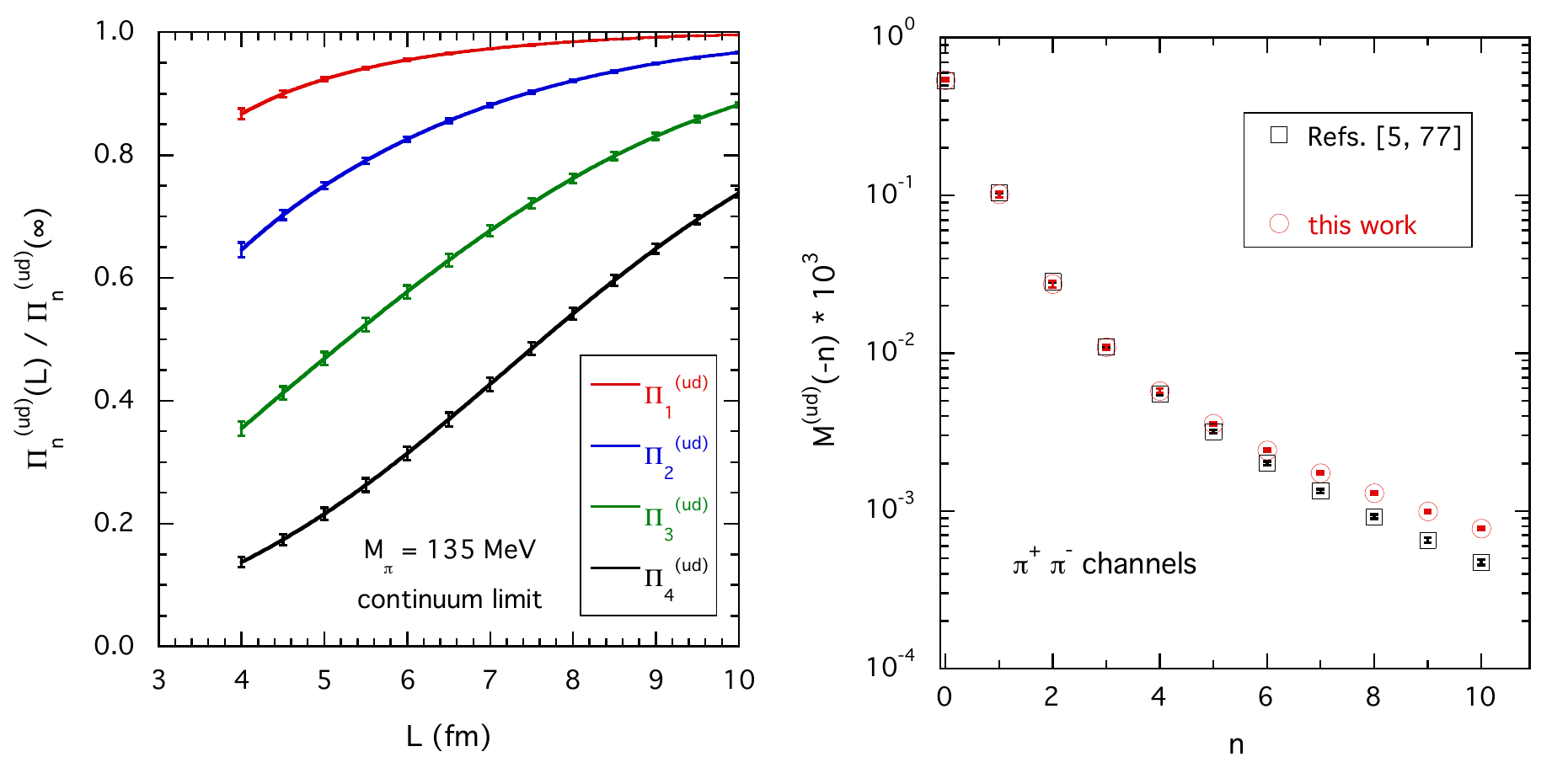}}}
\vspace{-0.25cm}
\caption{\it \small Left panel: ratio of the time moments (\ref{eq:Pin}) evaluated at finite lattice size $L$ and in the infinite volume limit using our ``dual + $\pi \pi$" analytic representation of the light-quark vector correlator taken in the continuum limit and at the physical pion point. Right panel: values of the first eleven moments (\ref{eq:Mn}) evaluated at the physical point using the $\pi \pi$ contribution (\ref{eq:V2pi_vol}) in the infinite volume limit (red circles), compared with the results of the dispersive analysis of the experimental cross section for the $e^+ e^- \to \pi^+ \pi^-$ channels of Ref.~\cite{Keshavarzi:2018mgv}. Courtesy of the authors of Ref.~\cite{KNT}.\hspace*{\fill}}
\label{fig:moments}
\end{figure}
In the left panel of Fig.~\ref{fig:moments} we show the FVEs on the ratio of the lowest four moments $\Pi_1^{(ud)}$ - $\Pi_4^{(ud)}$ evaluated at finite lattice size $L$ and in the infinite volume limit.
Thanks to the correlations between the numerator and the denominator the results for such ratios turn out to be very precise.
The impact of FVEs is sizeable and increases significantly as the order of the moment increases.
In the case of $\Pi_2^{(ud)}$ the use of a lattice size $L \sim 10$ fm still requires a finite volume correction equal to $\simeq 3 - 4 \%$. 

In the case of higher moments $\Pi_n^{(ud)}$ with $n > 2$ a reliable determination requires to reach very large time distances, i.e.~$t \gtrsim 4 $ fm.
This represents a stringent test for the large time-distance tail of the vector correlator $V^{(ud)}(t)$ evaluated with our analytic representation. 
The authors of Ref.~\cite{Keshavarzi:2018mgv} have kindly supplied us with the first eleven moments corresponding to the experimental cross section for the $e^+ e^- \to \pi^+ \pi^-$ channels only~\cite{KNT}.
The definition of the moments is slightly different from Eq.~(\ref{eq:Pin}) and follows the notation of Ref.~\cite{Charles:2017snx}, namely
\be
    \label{eq:Mn}
    {\cal{M}}^{(ud)}(-n) \equiv 4 \pi \alpha_{em} (-)^n (4 M_\pi^2)^{n + 1} ~ \frac{5}{9} ~ \Pi_{n + 1}^{(ud)} ~ .
\ee

We have evaluated Eq.~(\ref{eq:Mn}) using the $\pi \pi$ contribution  (\ref{eq:V2pi_vol}) in the infinite volume and continuum limits at the physical pion point.
The results are shown in Table~\ref{tab:moments} and in the right panel of Fig.~\ref{fig:moments} and they are compared with the dispersive values from Refs.~\cite{Keshavarzi:2018mgv,KNT}.
\begin{table}[hbt!]
\begin{center}
\renewcommand{\arraystretch}{1.10}
\begin{tabular}{||c||c||c||}
\hline
${\cal{M}}(-n) \cdot 10^3$ & dispersive $\pi^+ \pi^-$ & $\pi \pi$ representation \\ 
 Eq.~(\ref{eq:Mn})             & Ref.~\cite{KNT} & this work \\
\hline \hline
0 & 0.5336 ~ (21) & 0.5394 ~ (122) \\
\hline \hline
1 & 0.1046 ~~ (6) & 0.1021 ~ (45) \\
\hline
2 & 0.0285 ~~ (3) & 0.0274 ~ (13) \\
\hline \hline
3 & 0.01099 ~ (17) & 0.01091 ~ (42) \\
\hline
4 & 0.00549 ~ (11) & 0.00576 ~ (17) \\
\hline \hline
5 & 0.003183 ~ (75) & 0.003569 ~ (89) \\
\hline
6 & 0.002009 ~ (53) & 0.002420 ~ (54) \\
\hline
7 & 0.001336 ~ (39) & 0.001737 ~ (36) \\
\hline
8 & 0.000921 ~ (29) & 0.001298 ~ (26) \\
\hline
9 & 0.000653 ~ (22) & 0.000995 ~ (19) \\
\hline
10 & 0.000472 ~ (17) & 0.000775 ~ (15) \\
\hline
\end{tabular}
\renewcommand{\arraystretch}{1.0}
\end{center}
\vspace{-0.25cm}
\caption{\it \small Values of the first eleven moments (\ref{eq:Mn}) from the dispersive analysis of the experimental cross section for the $e^+ e^- \to \pi^+ \pi^-$ channels~\cite{KNT} and the corresponding ones evaluated at the physical point using the $\pi \pi$ contribution (\ref{eq:V2pi_vol}) in the infinite volume limit.\hspace*{\fill}}
\label{tab:moments}
\end{table}

Our results agree within the errors with the dispersive ones for $n \leq 4$, while they overestimate the dispersive moments at higher values of $n$.
It should be kept in mind that the values of Ref.~\cite{KNT} include the contributions of $u$ and $d$-quark disconnected diagrams as well as also IB effects.
Thus, the differences visible in Table~\ref{tab:moments} and in the right panel of Fig.~\ref{fig:moments} may be ascribed (at least partially) to the fact that the above contributions are missing in our calculations. 
Nevertheless, the good consistency visible (at least) for $n \leq 4$ indicates in our opinion that the large time-distance behavior of the vector correlator $V^{(ud)}(t)$ can be reliably evaluated using our analytic representation (at least for time distances currently accessible on the lattice).

Recently~\cite{Calame:2015fva,Abbiendi:2016xup} it has been proposed to determine $a_\mu^{\rm HVP}$ by measuring the running of $\alpha_{em}(q^2)$ for space-like values of the squared four-momentum transfer $q^2$ using a muon beam on a fixed electron target. 
The method is based on the following alternative formula for calculating $a_\mu^{\rm HVP}$~\cite{Lautrup:1971jf}:
 \be
    a_\mu^{\rm HVP} = \frac{\alpha_{em}}{\pi} \int_0^1 dx (1 - x) \Delta \alpha_{em}^{\rm HVP}[q^2(x)] ~ ,
    \label{eq:amu_muon}
 \ee
where $\Delta \alpha_{em}^{\rm HVP}(q^2)$ is the hadronic contribution to the running of $\alpha_{em}(q^2)$ evaluated at
 \be
     q^2(x) \equiv - \frac{x^2}{1-x} m_\mu^2 ~ .
     \label{eq:q2_muon}
 \ee
The quantity $\Delta \alpha_{em}^{\rm HVP}(q^2)$ can be extracted from the $q^2$-dependence of the $\mu e \to \mu e$ cross section data after the subtraction of the leptonic and weak contributions~\cite{Calame:2015fva,Abbiendi:2016xup}.
For the proposed MUonE experiment exploiting the muon beam at the CERN North Area~\cite{CarloniCalame:2017jfa} the region $x \in [0.93, 1]$ in Eq.~(\ref{eq:amu_muon}) cannot be reached and, therefore, the corresponding contribution
 \be
    [a_\mu^{\rm HVP}]_> \equiv \frac{\alpha_{em}}{\pi} \int_{\bar{x}}^1 dx (1 - x) \Delta \alpha_{em}^{\rm HVP}[q^2(x)] 
    \label{eq:amu_muon_>}
 \ee
with $\bar{x} = 0.93$ needs to be estimated using either $e^+ e^-$ data or lattice QCD simulations.
In terms of the Euclidean formula (\ref{eq:amu_t}) one has
 \be
      [a_\mu^{\rm HVP}]_> = 4 \alpha_{em}^2 \int_0^\infty dt ~ f_>(t) V(t) ~ ,
      \label{eq:amu_>}
 \ee
 where 
  \be
      f_>(t) \equiv \frac{4}{m_\mu^2} \int_{\bar{z}}^\infty dz ~ \frac{1}{\sqrt{4 + z^2}} ~ \left( \frac{\sqrt{4 + z^2} - z}{\sqrt{4 +z^2} +z} \right)^2 
                           \left[ \frac{\mbox{cos}(z\,m_\mu t) -1}{z^2} + \frac{1}{2} m_\mu^2 t^2 \right] 
      \label{eq:ftilde_>}
  \ee
with $\bar{z} = \bar{x} / \sqrt{1 - \bar{x}} \simeq 3.5$.

Using the analytical representation (\ref{eq:dual+2pi}) of the vector correlator $V_{ud}(t)$, evaluated  at the physical pion point in the continuum and infinite volume limits, the light-quark (connected) contribution $[a_\mu^{\rm HVP}]_>(ud)$ is found to be equal to
 \be
     [a_\mu^{\rm HVP}]_>(ud) = (81.2 \pm 1.7) \cdot 10^{-10} ~ .
     \label{eq:amu_>_ud}
 \ee
While the estimate of $[a_\mu^{\rm HVP}]_>$ requires also the addition of the contributions of the connected strange and charm quark terms as well as of disconnected and IB effects, our finding (\ref{eq:amu_>_ud}) indicates that the uncertainty of $[a_\mu^{\rm HVP}]_>$ should be of the order of $\simeq 2 \cdot 10^{-10}$.
Such a value is close to the statistical uncertainty ($\simeq 0.3 \%$) expected in the MUonE experiment for the contribution $[a_\mu^{\rm HVP}]_< \equiv [a_\mu^{\rm HVP}] - [a_\mu^{\rm HVP}]_>$ after two years of data taking at the CERN North Area~\cite{CarloniCalame:2017jfa}.

\section{Conclusions}
\label{sec:conclusions}

We have presented a lattice calculation of the leading HVP contribution of the light $u$- and $d$-quarks to the anomalous magnetic moment of the muon, $a_\mu^{\rm HVP}(ud)$.
The gauge configurations generated by ETMC with $N_f = 2+1+1$ dynamical quarks at three values of the lattice spacing ($a \simeq 0.062, 0.082, 0.089$ fm) and with pion masses in the range $M_\pi \simeq 210 - 450$ MeV have been adopted. 

Thanks to several lattices at fixed values of the light-quark mass and scale but with different sizes, we have performed a careful investigation of FVEs, which represent one of main source of uncertainty in modern lattice calculations of $a_\mu^{\rm HVP}(ud)$.
In order to remove them we have developed an analytic representation of the vector correlator and applied it to describe the lattice data for time distances larger than $\simeq 0.2$ fm.
The analytic representation is based on quark-hadron duality at small time distances and on the two-pion contributions in a finite box at larger time distances, assuming the GS parametrization~\cite{Gounaris:1968mw} for the timelike pion form factor.
Our estimate of FVEs takes into account the resonant interaction in the two-pion system at variance with the ChPT prediction at NLO~\cite{Aubin:2015rzx}.

After removing FVEs we have extrapolated the corrected lattice data to the physical pion point and to the continuum limit taking into account the chiral logs predicted by ChPT, obtaining
 \be
      a_\mu^{\rm HVP}(ud) = 619.0 ~ (17.8) \cdot 10^{-10} ~ ,
 \ee  
which is consistent with recent lattice results available in the literature~\cite{Chakraborty:2016mwy,DellaMorte:2017dyu,Borsanyi:2017zdw,Blum:2018mom}.

Adding the contribution of strange and charm quarks, $a_\mu^{\rm HVP}(s) = 53.1 ~ (2.5) \cdot 10^{-10}$ and $a_\mu^{\rm HVP}(c) = 14.75 ~ (0.56) \cdot 10^{-10}$ determined by ETMC in Ref.~\cite{Giusti:2017jof}, and an estimate of the IB corrections $a_\mu^{\rm HVP}(IB) = 8 ~ (5) \cdot 10^{-10}$ and of the quark disconnected diagrams $a_\mu^{\rm HVP}(disconn.) = -12 ~ (4) \cdot 10^{-10}$, obtained using the findings of Refs.~\cite{Borsanyi:2017zdw} and \cite{Blum:2018mom}, we get
 \be
     a_\mu^{\rm HVP}(udsc) = 683 ~ (19) \cdot 10^{-10} ~ ,
 \ee
 which agrees nicely with the recent results $a_\mu^{\rm HVP} = 688.07 ~ (4.14) \cdot 10^{-10}$~\cite{Jegerlehner:2017lbd}, $a_\mu^{\rm HVP} = 693.10 ~ (3.40) \cdot 10^{-10}$~\cite{Davier:2017zfy} and $a_\mu^{\rm HVP} = 693.27 ~ (2.46) \cdot 10^{-10}$~\cite{Keshavarzi:2018mgv}, based on dispersive analyses of the experimental cross section data for $e^+ e^-$ annihilation into hadrons. 

Using our analytic representation of the light-quark vector correlator, taken at the physical pion mass in the continuum and infinite volume limits, we have provided the slope and curvature of the polarization function, $ \Pi_1^{(ud)} = 0.1642 ~ (33) ~ \mbox{GeV}^{-2}$ and $\Pi_2^{(ud)} = - 0.383 ~ (16) ~ \mbox{GeV}^{-4}$, which have been compared with the corresponding lattice results of Ref.~\cite{Borsanyi:2016lpl}.
We have also evaluated the first eleven moments of the polarization function and compared them with the results of the dispersive analysis of the $\pi^+ \pi^-$ channels of Refs.~\cite{Keshavarzi:2018mgv,KNT}.
Finally, we have estimated the light-quark contribution to the missing part of $a_\mu^{\rm HVP}$ not covered in the MUonE experiment~\cite{Calame:2015fva,Abbiendi:2016xup} (see Eq.~(\ref{eq:amu_>_ud})).

New simulations with $N_f = 2 + 1 + 1$ dynamical quarks close to the physical pion point~\cite{Alexandrou:2018egz}, the evaluation of quark disconnected diagrams and of the IB corrections \cite{Giusti:2018vrc} are in progress by ETMC.
This will be crucial for improving the determination of the HVP contribution $a_\mu^{\rm HVP}(udsc)$ and for addressing the muon anomaly from first principles.

\section*{Acknowledgments}
We warmly thank our colleagues R.~Frezzotti, V.~Lubicz, G.~Martinelli and N.~Tantalo for enjoyable discussions and useful comments.
We thank S.~Bacchio and B.~Kostrezwa for supplying us with the DD$\alpha$AMG library~\cite{Alexandrou:2016izb} integrated in the tmLQCD software package~\cite{Jansen:2009xp,Abdel-Rehim:2013wba,Deuzeman:2013xaa}.
A.~Keshavarzi, D.~Nomura and T.~Teubner are very gratefully acknowledged for having provided the values of the first eleven moments of the polarization function~\cite{KNT} corresponding to the dispersive analysis of the $\pi^+ \pi^-$ channels of Ref.~\cite{Keshavarzi:2018mgv}.
We gratefully acknowledge the CPU time provided by CINECA under the initiative INFN-LQCD123 on the KNL system Marconi at CINECA (Italy).

\end{document}